# Establishing a common database of ice experiments and using machine learning to understand and predict ice behavior

Leon Kellner[a,1], Merten Stender[b], Rüdiger U. Franz von Bock und Polach[a], Hauke Herrnring[a], Sören Ehlers[a], Norbert Hoffmann[b,c], Knut V. Høyland[d]

[a] Institute for Ship Structural Design and Analysis, Hamburg University of Technology, Germany
[b] Dynamics Group, Hamburg University of Technology, Germany
[c] Department of Mechanical Engineering, Imperial College London, England
[d] Sustainable Arctic Marine and Coastal Technology (SAMCoT), Centre for Research-based Innovation (CRI), Norwegian University of Science and Technology, Trondheim, Norway

## Abstract

Ice material models often limit the accuracy of ice related simulations. The reasons for this are manifold, e.g. complex ice properties. One issue is linking experimental data to ice material modeling, where the aim is to identify patterns in the data that can be used by the models. However, numerous parameters that influence ice behavior lead to large, high dimensional data sets which are often fragmented. Handling the data manually becomes impractical. Machine learning and statistical tools are applied to identify how parameters, such as temperature, influence peak stress and ice behavior. To enable the analysis, a common and small scale experimental database is established.

# 1. Introduction

## 1.1. Motivation

Material modeling of ice still remains a challenge due to its inherent complexity as well as gaps in experimental data. This research focuses on linking experimental data to ice models and underlying ice mechanics. A large database of existing ice experiments is established and analyzed with machine

---

[1] Corresponding author. Mail: leon.kellner@tuhh.de. Phone: +49 42878 6085. Address: Institute for Ship Structural Design and Analysis. Am Schwarzenberg Campus 4 C. 21073 Hamburg. Germany

learning and statistical tools to identify patterns. The aim is to support material modeling regarding decisions on which and how to include parameters, such as temperature or strain rate, in models for ice related simulations. Moreover, the addressed question is whether these patterns agree with the state-of-the-art understanding of ice mechanics.

Generally, ice mechanics are investigated on a broad range of scales and applications, from single crystals and small laboratory experiments with specimens of the size of centimeters to glaciers and ice sheets, see e.g. (Montagnat et al., 2014; Schulson, 2015). In this research, the focus is on experimental data and the identification of dominant features regarding ice behavior. This is seen as a first step towards material models for ice related simulations of different applications.

The complexity of ice is illustrated in ice-structure interaction (ISI) processes. The forces acting between the structure and the ice fluctuate considerably both in space and time (Ralph and Jordaan, 2017; Yue et al., 2009). Most of the global load is transmitted though small regions of the total interaction area, called high pressure zones. Within these zones, ice is highly confined and can transmit high local loads. Lower confinement, on the other hand, limits the ability of ice to transmit loads. Hence, further away from the central area, less confined ice tends to fail under compression. This leads to fracture, spalling and extrusion of material, see (Jordaan, 2001) for a review of ISI.

This behavior is a result of the material properties of ice. The characteristics of ice change depending on numerous parameters such as strain rate (Petrovic, 2003), grain size (Batto and Schulson, 1993), or confinement, e.g. (Renshaw et al., 2014). The occurring phenomena cover viscous ductile or brittle behavior, hardening and softening, dynamic recrystallization as well as size dependent fracture (Schulson and Duval, 2009; Timco and Weeks, 2010). Consequently, ice material models that reflect this complexity are needed for accurate ice interaction simulations. However, in contrast to well established steel material models for the ship or offshore structure, there is no standardized material model available for ice that can capture all effects during the interaction process.

A first step towards complete material models for e.g. ice interaction simulations is the development of material models suitable for the simulations of small scale experiments. This comes with the benefit of a controlled environment and known material properties. Nevertheless, the difficulties developing material models for small scale simulations are already manifold.

To begin with, there are gaps in experimental data. For example, little data exists on shear and tensile strength of ice (Timco and Weeks, 2010). Also, a lot of data exists for ice under uniaxial compression, but much less for bi- and triaxial tests under high confinement. Yet, in reality, ice is often under a multi-axial state of stress (Schulson and Duval, 2009). Regarding ice mechanics, many topics are under investigation, for instance how damage and recrystallization alter ice properties (Snyder, 2015), how ice fails under high triaxial stress states (Renshaw and Schulson, 2017), the scaling nature of tensile fracture (Dempsey et al., 2018) or the physics of ice-ice friction (Schulson, 2018). Additionally, some areas remain subject of ongoing discussion, for example the scaling of small scale and model ice tests (Palmer and Dempsey, 2009; Palmer and Croasdale, 2013; von Bock und Polach and Molyneux, 2017), or ice-induced vibrations (Määttänen, 2015).

Also, much data exists, see e.g. (Timco and Weeks, 2010). Yet, to the authors' knowledge there is no general purpose and publicly available experimental database, though desirable and for instance recommended by (Ehlers et al., 2018). Benefits are for instance that the analysis of large data sets can make the results less susceptible to variability, or the possibility to investigate distributions of results.

When data is collected and combined, it is often limited to specific measurements, e.g. to investigating the relation between flexural strength and brine volume (Timco and O'Brien, 1994). On the other hand, general purpose data sets need to be high dimensional to capture the complex ice properties and all potentially influencing parameters. As a result, such data sets can become too large to handle manually. Lastly, it is not straightforward to handle missing measurements.

The ability of machine learning and statistical methods to identify patterns in data is particularly useful under such circumstances, see e.g. (Chicco, 2017; Larrañaga et al., 2006). Yet these methods are applied rarely or not at all in ice mechanics.

Here, machine learning and complementary statistical methods are used to derive relationships from the data and draw conclusions that can be used in ice material modeling. More specifically, the objective is to identify how parameters, such as temperature or grain size, influence peak stress and ice behavior. It is investigated whether the results from the analysis, based on a large data set, agree with commonly observed experimental relationships. To enable the analysis, a common and general purpose small scale experimental database is established, to be used and extended by everyone.

## 1.2. Overview and methodology

A graphical abstract is given in Figure 1. In the first part of this paper, Section 2, the focus is on establishing the database. A literature review is done to identify all parameters (i.e. features in data analysis) that influence the outcome of ice experiments. In this regard, the collection of data is considered general purpose. However, for the current database some limitations apply. The data consists of small scale compression and tensile test data, because a lot of data is available and the process of systematically gathering it is straightforward. Other types of tests, e.g. creep or prestrain, are not included, because the systematic collection is more elaborate due to an increased number of influencing features, e.g. prestrain level and prestrain strain rate.

The data analysis is described in the second part, Sections 3 and 4. It consists of two steps; filtering the data to obtain method specific sub data sets and applying the machine learning and statistical methods to those sub data sets.

With respect to peak stress, correlation and model reduction techniques are applied to the data. A common method to quantify univariate correlation is the Pearson correlation coefficient (PCC). For instance, it is used to investigate relationships between symptoms for post-traumatic stress disorder and post-traumatic growth (Liu et al., 2017). Here it is applied to correlating features to peak stress and thereby identifying the most important features.

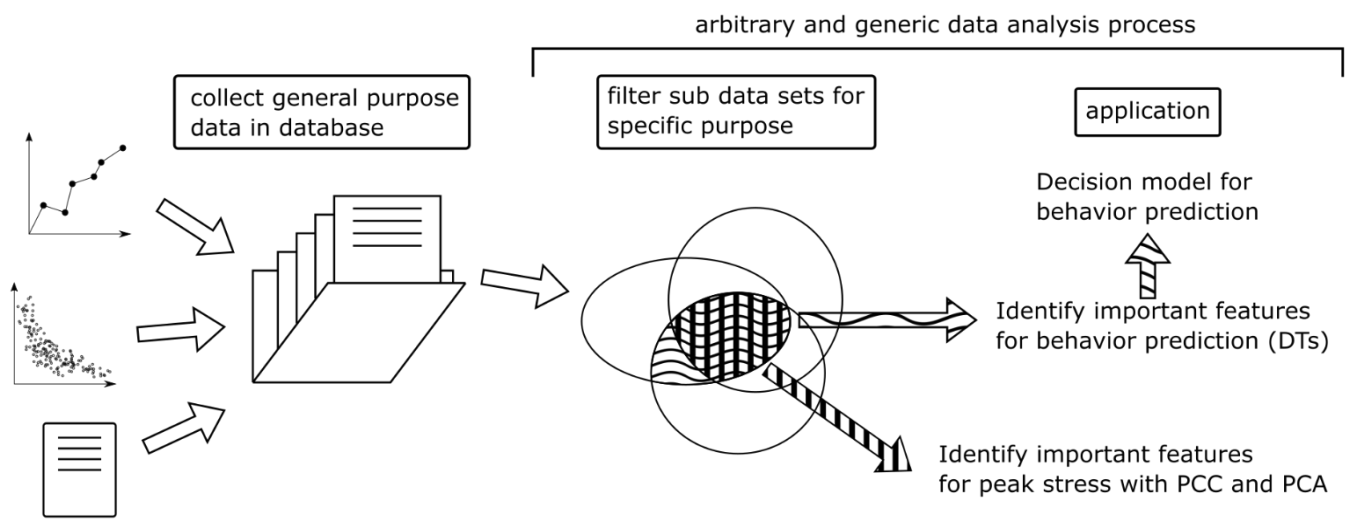

**Figure 1 - Graphical abstract**

Since there exist many potential influencing features in experiments, e.g. temperature or grain size, model reduction is applied to identify irrelevant or redundant features. A typical tool for model reduction is the principal component analysis (PCA). It aims to reduce data dimensionality by transforming a set of multivariately correlated variables to a smaller set of uncorrelated variables ('principal components'). The smaller set still explains most of the information of the original set while better capturing its variability. It is commonly used to examine the contribution of variables to variance in data and pattern recognition, e.g. (Keeley and McDonald, 2015; Mantovani et al., 2012).

Because ice changes its behavior under different conditions, one aim is also to investigate the predictive value of features with regard to the type of ice behavior, e.g. ductile, splitting etc. This is a supervised classification task, and various machine learning models can be applied. Preliminary runs with different models point at using decision trees (DT) for this. They are a widely used classification tool and work for both categorical and numerical data. In addition to that, results are easy to interpret and it is straightforward to handle missing data via surrogate splits. In principle, they recursively partition the data set into subsets to form conditional statements about the data. These decision rules help find the most decisive features leading to a class of observations.

There is a wide range of applications for DTs, for example the prediction of protein types (Sankari and Manimegalai, 2017) or the identification of risk factors associated with metabolic syndrome (Tayefi et al., 2018). The DTs are applied in order to rank the most important features for ice behavior prediction. Based on this ranking, the input data for training final DTs is chosen. Lastly, two models are given to predict global brittle or ductile behavior in saltwater and freshwater ice experiments.

Generally, the choice of the described methods is guided by interpretability and straightforward application to models, which doesn't favor other options such as dimensional analysis or different machine learning methods e.g. artificial neural networks or support vector machines (Chicco, 2017; Palmer, 2008).

Lastly, in Sections 4 and 5 the results are discussed and compared to the commonly agreed ice mechanics relationships described in Section 2.1.

## 2. Database

Below, the database is presented. To begin with, factors that are thought to influence the peak stress and behavior of ice are reviewed. Then, the methodology of data collection is outlined.

### 2.1. What to include

To enable decisions about what to include in the database, the response of ice to stress or load under different influencing factors (features) is investigated. Here, the response comprises two aspects. First, the peak stress the specimen could sustain, i.e. its strength. Second, the qualitative type of behavior such as ductile or brittle and, where possible, more specific sub-types, e.g. Coulombic faulting.

Influencing features can be either intrinsic properties, e.g. grain size, or external factors such as confinement or stress state. The assessment is done for small scale experimental data without going into detail. The causes for specific behavior are not described here. For an in-depth description of ice as a material and its behavior the reader is referred to e.g. (Fletcher, 1970; Hobbs, 2010; Palmer and Croasdale, 2013; Sanderson, 1988; Schulson and Duval, 2009).

A systematic overview of influential features should start with the history of the material. Ice usually contains flaws. Inclusions, e.g. gas, and damage can exist depending on formation, age and precursory stress (Stone et al., 1997; Timco and Weeks, 2010). Its history also influences other intrinsic properties such as ice type, grain size, porosity (gas and brine), temperature and salinity. These, in turn, affect the response to applied stress.

One way to see this is by looking at strength, or peak stress, respectively. For example, uniaxial tensile strength appears to be sensitive to specimen size, grain size and brine volume and less sensitive to temperature (Hawkes and Mellor, 1972; Schulson et al., 1984; Schulson, 1999). Compressive strength appears to be insensitive to size (Schulson and Duval, 2009), but depends on grain size (Schulson, 1990), porosity and temperature (Moslet, 2007), and ice type (Kuehn and Schulson, 1994). Prestrain or damage also affects the compressive strength and behavior type (Snyder et al., 2016).

Besides intrinsic properties, external factors must be considered. They are typically connected to test conditions. However, from the material's perspective it comes down to stress state, strain rate and strain. First, ice under uniaxial tension is considered. Under this condition, ice tends to fail suddenly under small strains ($\sim 0.01 - 0.1\,\%$), mostly in a brittle manner and independent of strain rate (Hawkes and Mellor, 1972; Schulson et al., 1984; Schulson, 1999). To the authors' knowledge, there is no data on tests under bi- or multiaxial tensile stress states.

In contrast to tension, the behavior of ice under compressive stress states depends on strain rate (Ince et al., 2016; Timco and Weeks, 2010). Different combinations of stress state and strain rate result in changes of behavior. Under uniaxial compression, both brittle, e.g. axial split, and ductile failure can be observed. With increasing confinement, that is bi- or triaxial stress states, other behavior types occur, for instance Coulombic- and plastic faulting (Renshaw et al., 2014). In practice, ice is generally under a triaxial state of stress.

Next, the maximum or failure strain is discussed. Brittle failure under compression is linked to small failure strains $\sim \epsilon \leq 0.5\,\%$ and higher strain rates $\sim \dot{\epsilon} \geq 10^{-3}$. For ductile behavior, strains can be larger. Under very low strain rates, i.e. $\sim \dot{\epsilon} \leq 10^{-5}$, ice can even sustain strains of more than one

(Schulson and Duval, 2009). However, such low strain rates are more related to flow in glaciers and ice sheets and less important for ice interaction scenarios.

One aim of this research is to evaluate if the above described experimental correlations can be validated through statistical and machine learning analysis of the large database. Despite this being done regarding peak stress and ice behavior, the same approach could be applied to other correlations in ice mechanics.

## 2.2. Features and observations

The database comprises about 3000 observations. Each observation represents one test run[2] with several measurements or features, respectively, e.g. peak stress, porosity, temperature etc. This includes uni-, bi- and triaxial compression tests and uniaxial tensile tests under different conditions. The composition and the sources are given in Table 1. The decision about which features to include is based on the above review. The database is in table format, rows correspond to observations and columns correspond to the features described below and their units. Generally, the aim was not to collect all available data, but rather to initiate the database and obtain enough observations to demonstrate the analysis methods.

Table 1 - Overview of database

| Total number of entries | | 2939 | |
|---|---|---|---|
| **Of which:** | | | |
| **Type of experiment** | | **Behavior classification** | |
| uniaxial compression | 2303 | brittle | 979 |
| biaxial compression | 228 | c-fault | 33 |
| triaxial compression | 354 | p-fault | 33 |
| uniaxial tension | 53 | ductile | 1329 |
| shear[3] | 1 | transitional | 82 |
| | | other or undefined | 549 |
| **Type of water used to make the ice** | | | |
| freshwater | 729 | | |
| saltwater | 2211 | | |
| undefined | 3 | | |
| **Sources:** | (Arakawa and Maeno, 1997; Currier and Schulson, 1982; Golding et al., 2010, 2012; Golding et al., 2014; Gratz and Schulson, 1997; Häusler, 1981; Hawkes and Mellor, 1972; Haynes, 1978; Haynes and Mellor, 1977; Iliescu and Schulson, 2004; Jones, 1982, 1997; Mellor and Cole, 1982; Mizuno, 1998; Nadreau et al., 1991; Richter-Menge and Jones, 1993; Rist and Murrell, 1994; Schulson, 1990, | | |

---

[2] Sometimes only average values from several tests are published. In that case those values are taken as a single observation, although they reflect more than one test.

[3] Using only one value for shear experiments follows Timco and Weeks, who write "many […] results were generated using test techniques which impose unrealistic (and unknown) normal stresses on the failure plane." and give limit values from "the more reliable tests" which we simply averaged and took in as one 'example' shear value (Timco and Weeks, 2010).

| | 1999; Schulson et al., 2006; Schulson and Buck, 1995; Schulson and Gratz, 1999; Smith and Schulson, 1993; Strub-Klein, 2017; Strub-Klein and Sudom, 2012; Timco and Frederking, 1984; Timco and Weeks, 2010; Weiss and Schulson, 1995; Zhang et al., 2011; ZhiJun et al., 2011) |
|---|---|
| | Additionally, data from the UNIS project from 2004 to 2011 is included: (Høyland, 2007; Moslet, 2007; Strub-Klein and Sudom, 2012) |

In the following, the features included in the database are listed and discussed if required. Some general remarks are given afterwards.

1. Type of test: shear, tensile, uniaxial/biaxial/triaxial compression

Uniaxial and biaxial compression indicate no stresses in two or one directions, respectfully.

2. Global principal stresses $\sigma_1, \sigma_2, \sigma_3$ in MPa

Includes the global principal stresses as given by the experimental setup at the time of peak stress. The stress distribution in the specimen is expected to be heterogeneous due to the crystal structure. As a result, local stress states can vary both in direction and magnitude (Grennerat et al., 2012).

3. Global peak stress $\sigma_p$ in MPa

Here, peak stress is defined as the global maximum stress the specimen could sustain during the experiment, usually in the direction of the biggest principal stress, e.g. for uniaxial tensile tests it is the tensile strength, for uniaxial compressive tests it is compressive strength. For brittle failure, peak stress is also failure stress. For ductile failure, this is given as the maximum of the stress-strain or stress-displacement curve during the experiment. Like principal stresses, local stress may be higher and not oriented in the same direction as the global peak stress.

4. Terminology for peak stress as originally described by the author

Different synonyms for peak stress are used by researchers. Peak stress is also termed critical stress, maximum stress or strength etc. In some cases it is also called yield stress or yield strength which implies material plasticity, although it is defined as peak stress. If one of these synonyms is used, it is indicated in this column.

5. Homogenized qualitative behavior.

Here, the original behavior description by the author is homogenized as far as possible. The aim is to enable better analysis of the data based on an agreed set of behavior categories. So far, this set includes the following:

- Brittle
    - C-Fault
    - P-Fault
    - Brittle shear
    - Brittle tensile
- Ductile

- Transition

A corresponding type from the list is chosen according to the original description or if applicable, from the characteristics of the given stress - strain or load - time curves. A smooth trajectory indicates ductile behavior whereas for brittle behavior the curves show a sharp peak followed by a sudden drop (Schulson, 1999; Snyder et al., 2016). C- and P-faulting are two distinct modes of brittle-like shear faulting under compression, for instance described in (Golding et al., 2010; Renshaw et al., 2014). Brittle shear and brittle tensile indicates failure in pure shear, e.g. asymmetric four-point bending, and pure tensile experiments. The term transition indicates intermediate behavior between ductile and brittle. It is only used when explicitly mentioned by the original authors. Some exemplary behavior categories are shown in Figure 2.

6. Limit strain, $\epsilon$

This indicates global limit strain. It is usually calculated with the displacement and the initial specimen length(s). If published, limit strain is included in all directions. Local strain fields are usually heterogeneous and should not be confused with the global strain.

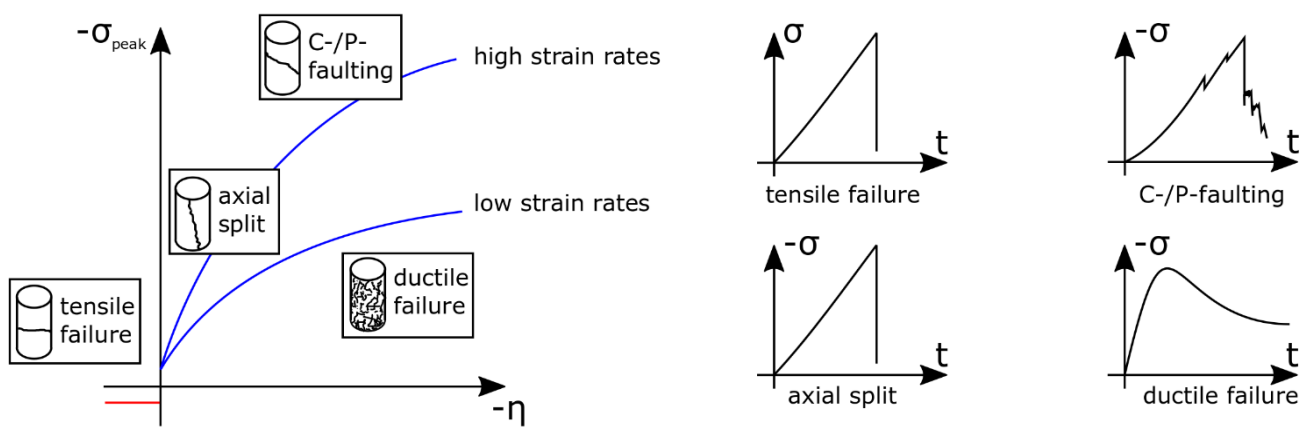

Figure 2 - Exemplary behavior categories in dependence of triaxiality and strain rate (left) and exemplary stress time curves (right).

7. Type of limit strain

This indicates whether the global limit strain is given at peak or failure stress. Additionally, in the future, this should include information as to how the strain was obtained. Several factors influence limit strain (and also strain rate) for instance interface effects or the stiffness of the test setup.

8. Global strain rate, $\dot{\epsilon}$ in $\frac{1}{s}$

This indicates global strain rate as given by the original author. Typically, the global strain rate is calculated based on global strain at peak stress. Like principal stresses and local strains, local strain rates can vary significantly (Grennerat et al., 2012).

9. Ice temperature in °C

For small specimens, this is usually given as the room temperature during storage and experiment. The temperature can also vary within the specimen, but this is expected to be less pronounced for small scale tests.

10. Grain size in mm

This is defined as the average grain size, which implies isotropic material and is usually only given for granular ice. However, it should be noted that the isotropy assumption is not always valid, depending on the ratio of specimen size to grain size (Dempsey et al., 1999). The grain size may also be given for columnar ice. In this case, since column diameters tend to vary, it must be clear where and how the grain size was measured. To the authors' knowledge, there is no standardized way to measure and calculate average grain size yet. Differences between measuring procedures are known, see e.g. (Lehto et al., 2014; Roebuck et al., 2004), and possibly influence the data.

11. Total porosity (brine and air) in %
12. Bulk salinity of the ice in ‰
13. Type of ice, e.g. columnar or granular/polycrystalline
14. Type of water used to make the ice

The following types comprise most of the water types used in experiments:

- Fresh water [f]
- Salt water [s]
- Tap water [t]
- Distilled/degassed/deionized water [d]

They are also sometimes combined, e.g. distilled tap water without degassing. This point is important for comparison. More information on water type is given in the comment column.

15. Columnar loading

For columnar ice it is documented whether it was loaded across, along or at a 45° angle to the columns. This influences its response to loading, see e.g. (Timco and Weeks, 2010).

16. Hydrostatic pressure, $\sigma_h$, in MPa

To compute triaxiality (see feature 18), the hydrostatic pressure, or mean stress, is calculated with the principal stresses as

$$\sigma_h = \frac{\sigma_1 + \sigma_2 + \sigma_3}{3} \qquad (1)$$

17. Equivalent, von Mises Stress, $\sigma_e$, in MPa

This is also used for computing the triaxiality with the principal stresses as input values and calculated as

$$\sigma_e = \frac{1}{\sqrt{2}}\sqrt{(\sigma_1 - \sigma_2)^2 + (\sigma_1 - \sigma_3)^2 + (\sigma_2 - \sigma_3)^2} \qquad (2)$$

**18.** Triaxiality, $\eta$

Triaxiality is a measure which relates hydrostatic, $\sigma_h$, to shear stress, which can be represented by the ratio of $\sigma_h$ and the von Mises stress, $\sigma_e$

$$\eta = \frac{\sigma_h}{\sigma_e} \tag{3}$$

Triaxiality is for instance used for plastic limit strain failure criteria in crashworthiness computations of metal sheets (Effelsberg et al., 2012). In Figure 3, some examples of two-dimensional stress states are given. A positive triaxiality value indicates a state of hydrostatic tension. Biaxial stress states are limited to values of $-\frac{2}{3} \leq \eta \leq \frac{2}{3}$, whereas $\eta$ is unlimited for three-dimensional stress states, e.g. for a three-dimensional equal principal stresses $\eta = \infty$. Pure shear stress states result in $\eta = 0$. Triaxiality values can be ambiguous, that is, different stress states can yield the same triaxiality value. Despite this ambiguity, triaxiality can be useful as it condenses information on the stress state into one value.

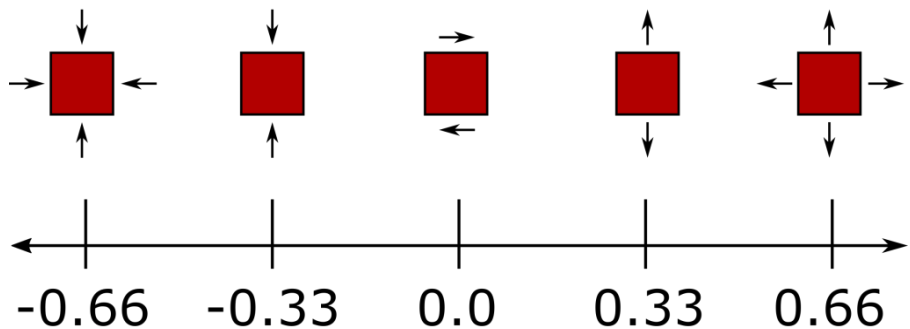

**Figure 3 – Exemplary triaxiality values for biaxial stress states. For three dimensional stress states, triaxiality values are unlimited.**

**19.** Specimen size

The dimensions of the specimen are documented as well as the volume. The reason for documenting this is the size dependence of failure under tension, see e.g. (Schulson, 1999).

## 2.3. Remarks

- This list is thought to comprise major influencing factors regarding peak stress and behavior types. It remains open to discussion and revision. Other researchers are invited to collaborate in using and extending the database.
- The data does not include any information on testing conditions yet, e.g. stiffness of test setup. The smaller the amount of available data for specific testing conditions, the greater the influence of the conditions. If a lot of different data sets exist for specific testing conditions, the influence of a single data set decreases. Nonetheless, testing conditions should be included in the future, particularly as long as no standardized test procedures exist.
- Currently, the database is not complete. For most features some data is missing, i.e. there are around 10-20% of missing entries. Large gaps are for example found for limit strain (90% missing) or porosity (43% missing), see also Table 7.
- The database is not balanced. Most of the data is concentrated around certain feature values. For example: 80% of observations are within a strain rate range $10^{-3} \leq \dot{\epsilon} \leq 10^{-2}$, 28% are within a temperature range $-11°\text{C} \leq T \leq -9°\text{C}$ and 95% of observations have triaxiality values of $-1 \leq \eta \leq 0$. Moreover, currently more salt- than freshwater ice data is included, see Table 1.
- The majority of data is from uniaxial tests. Uniaxial strength should be seen as an index strength rather than a material property (Strub-Klein, 2017).
- Most variables in the list only reflect global behavior of the specimen. Local stress, strain rate, temperature etc. can vary throughout the specimen. It should also be noted that, for now, only scalar data is included. No force-time or stress-strain curves are part of the data since they are rarely published to the full extent. Especially for the comparison to simulations such curves are useful and should be included in the future whenever possible. Lastly, qualitative data can leave room for interpretation, e.g. whether a force time curve indicates ductile or brittle behavior.

## 3. Using the data to understand and predict ice behavior

After establishing the database, the next step is the data analysis. In other words, the collected data is used for a specific purpose: to identify the most influential factors with respect to peak stress and the behavior type of ice and to build a model that predicts behavior type based on user input. The correlation analysis and the PCA focus on peak stress, whereas the goal of the decision trees the prediction of behavior type. Note that in the following, the factors or parameters are called features, which is in line with machine learning terminology.

### 3.1. Correlation analysis

Correlations of different features with the target quantity can be computed in numerous ways. A common approach is the visual inspection of a matrix of scatter plots correlating each input feature with the target variable. If both quantities are univariately related, distinct structures will form in the scatter plot. In case of a linearly positive relation, most data points will arrange along a diagonal. If no structure is observed there is no univariate relationship between both variables. The visual inspection allows to identify outliers and observe the distribution of individual variables.

To quantify the univariate and linear relationship between input features and the target variable Pearson (empirical) correlation coefficient $\rho$ can be employed, see e.g. (Gibbons, 1986). It is defined as

$$\rho = \frac{\sum(x_i - \bar{x})(y_i - \bar{y})}{\sqrt{\sum(x_i - \bar{x})^2 \cdot \sum(y_i - \bar{y})^2}} \qquad (4)$$

with $\overline{(\cdot)}$ being the mean of the respective set of observations. Correlation coefficients range from $-1$ to $1$ where $1$ indicates a perfect positive and $-1$ a perfect negative correlation of observations $x$ and $y$. Values close to $0$ indicate no linear univariate correlation. However, other, non-linear relationships between two variables might still exist.

Here, input features are separately correlated to the peak stress value for ductile and brittle material behavior, as well as for freshwater and saltwater ice. The aim is to identify which features are important to consider in any model that aims to predict peak stress. Moreover, it is investigated whether the most influential features change with respect to ice and behavior type.

## 3.2. Principal component analysis for the investigation of variability of data

The data matrix $X \in \mathbb{R}^{n \times m}$ contains $m$ measurements, so called features, for $n$ observations. Principal component analysis (PCA) allows reducing this high-dimensional data set to a low-dimensional representation that can reflect the data in fewer dimensions. The main assumption is that observations that do not have much variability cannot reflect much of the correlations in the database. In other words, the idea is that some measurements (resp. features) in $X$ are to a large degree irrelevant or redundant.

The principal components (PCs) are new base vectors that represent a different way of looking at the data, which gives more weight to high variability features. In contrast, the old base vectors correspond to standard axes, where every axis points in the direction of one feature, e.g. if moving along the temperature axis only temperature changes. The PCs are a combination of features and point in the direction of highest variance in the data point cloud. Thus, the data does not change, only the perspective. For instance, consider that temperature, peak stress, and grain size are recorded, but all experiments are done at the same temperature. Then, peak stress and grain size would account for all variance in the data, whereas temperature would not vary at all. Consequently, the first PC would be a combination of peak stress and grain size, but not temperature. The first PC would also account for 100% of variance in the data.

To be more specific, PCA is defined as the eigendecomposition of the co-variance matrix $(X^T X) \in \mathbb{R}^{m \times m}$. The set of eigenvalues $\lambda$ and eigenvectors ('loadings') $W$ $[m \times m]$ can be used to describe the data. The columns of $W$ are the principal components, which are ordered by the size of the corresponding eigenvalues. The size of the eigenvalues directly reflects the variance explained by the corresponding eigenvector / principal component. The full variance of the data set is given by the sum of the covariance matrix's eigenvalues $\lambda$

$$\Gamma = \sum_{j}^{n} \lambda_j \tag{5}$$

so that the variance explained by each eigenvalue (i.e. principal component) is

$$r_j = \lambda_j / \Gamma \tag{6}$$

The scores $T$ $[n \times m] = XW$ represent the data in a new basis spanned by the principal components, i.e. after the transformation by $W$. Since the columns of W are ordered by the value of the corresponding eigenvalues (i.e. the variance explained), dimensionality reduction can be achieved by truncating $W$. If we want to represent our data by the first r principal components, $W_r$ will be the first r columns of $W$. The reduced representation of the data is computed from the truncated basis $T_r = X W_r$.

However, the contribution of original features to the principal components does not explain the relevance of individual features for the prediction of some behavior type or peak stress value: A large variance does not necessarily relate to high importance for a prediction task. Also, multivariate, and possibly nonlinear relations of features might dominate ice behavior.

In short, in using the PCA we create features in a new, possibly dimension-reduced, space by linear transformation of the original feature space. If the physical interpretation is not key, the PCA results can be used to create a small number of new features as input quantities to some further analytics such as classification or regression learning models. Here, physical interpretability is to be maintained and the new features are not used further, e.g. as input for the decision trees. However, the PCA is applied to indicate important features with respect to the variance in the data. For further reading, see e.g. (Härdle and Simar, 2015).

### 3.3. Decision Trees for feature selection and behavior prediction

In the following, the focus shifts from peak stress to the behavior type of ice. The objective is to investigate the predictive value of all features, categorical and numerical, with regard to behavior type and identify the most important ones. Also, a model should be built to predict ice behavior based on experimental conditions as input by the user. Decision trees (DTs) are used to achieve this.

Decision trees (DT) are a machine learning approach for considering the multivariate nature of the data and nonlinear correlations of features. Hence, DTs can be used to classify or predict an observation with unknown label when trained sufficiently on labeled data. DTs are a class of supervised machine learning methods.

Again, some features will be more important for the modeling task than others – other input features can be irrelevant or redundant. Robustness and generalization of decision trees decrease with rising numbers of features. Therefore, a small number of input features is favored. Possibly, one could use the features derived from PCA as in Section 3.2. To maintain physical interpretability of the DT, we keep the original features. However, when applying feature selection before the classification one must make sure to cross-validate the complete process (Hastie et al., 2017). A detailed description of the DT application in this work can be found in the appendix, Section 8.2.

Here we use the decision tree itself as feature selection procedure by applying a specific warping method: the greedy backward elimination scheme; for a greedy selection theme see e.g. (Fulcher and Jones, 2014). Starting with the complete set of $N$ original features, all possible combinations of $n = (N-1)$ feature subsets are created by leaving out one feature. For each subset a classifier is trained, and the classification performance is evaluated. The subset forming the maximum classification quality is kept (i.e. the respective missing feature is eliminated). This process of eliminating features is iterated until a minimum performance is reached. All features that have been eliminated up to this point do not represent discriminative features and thus may be discarded. All remaining features form the highly discriminative feature subset which is physically interpretable (in contrast to the subset derived by PCA).

## 4. Results

### 4.1. Preliminaries

Prior to the analysis, the data is cleaned up. Measurements with unclear or undefined feature values are deleted or ignored. Some sub-categories are replaced with the higher-level category to keep it on a more general level. For instance, brittle-tensile is replaced with brittle, or distilled water is replaced with fresh water. Moreover, specimen geometry data is reduced to the largest dimension and volume. For cubes, the largest dimension is the longest side, for cylinders it is usually the length.

Three observations are excluded, because their triaxiality values are considered outliers[4], with $\eta = \{-507, -233, -142\}$, whereas for all other observations $\eta \geq -10$. Generally, temperatures, triaxialities and strain rates are set to NaN if outside the limits: $-100°\,\text{C} < T < 0°\,\text{C}$, $-10 < \eta < 1$ and $10^{-10} < \dot{\epsilon} < 10^{10}$. Shear data is excluded, because there is too little data available for a meaningful analysis.

Strain rates are logarithmized to base 10. Afterwards, 2939 observations remain. Further, method specific preprocessing is described in the respective sections. A general list of features and their completeness is given in the appendix, Table 7.

## 4.2. Correlation analysis

The correlation analysis is done in two steps. First, the visual inspection of a matrix of scatter plots correlating each input feature with the target variable, and second, the calculation of the Pearson correlation coefficients. The input data set comprises only continuous numerical, i.e. non-categorical, features. In Table 2, input features are given, along with the total number of observations and the completeness of observations, i.e. available data for that feature.

The plots correlating each feature with peak stress, grouped in ductile and brittle, are shown in Figure 4. The same plot grouped in freshwater and saltwater ice can be found in the appendix, Figure 15. Most of the data is scattered, but for some features the plots indicate a dependency. Linear relationships can be seen e.g. for the equivalent and hydrostatic stress, $\sigma_e$ and $\sigma_h$. A non-linear relationship would be indicated e.g. by exponential or sinusoidal relationships. This seems to be the case for the relation between porosity and peak stress, which would agree with literature, e.g. (Timco and Weeks, 2010). Nevertheless, the linear trend also approximates that increased porosity results in lower peak stresses. Other non-linear relationships are not obvious. Keeping this in mind, the linear Pearson correlation coefficient (PCC) can be used for further analysis.

---

[4] The data are taken from Häusler (1981). It is questionable that ice critically fails under such high confinement, since those triaxiality values indicate almost identical principal stresses; the differences between the principal stresses are $< 0.4\%$. This could also be close or below measurement accuracy.

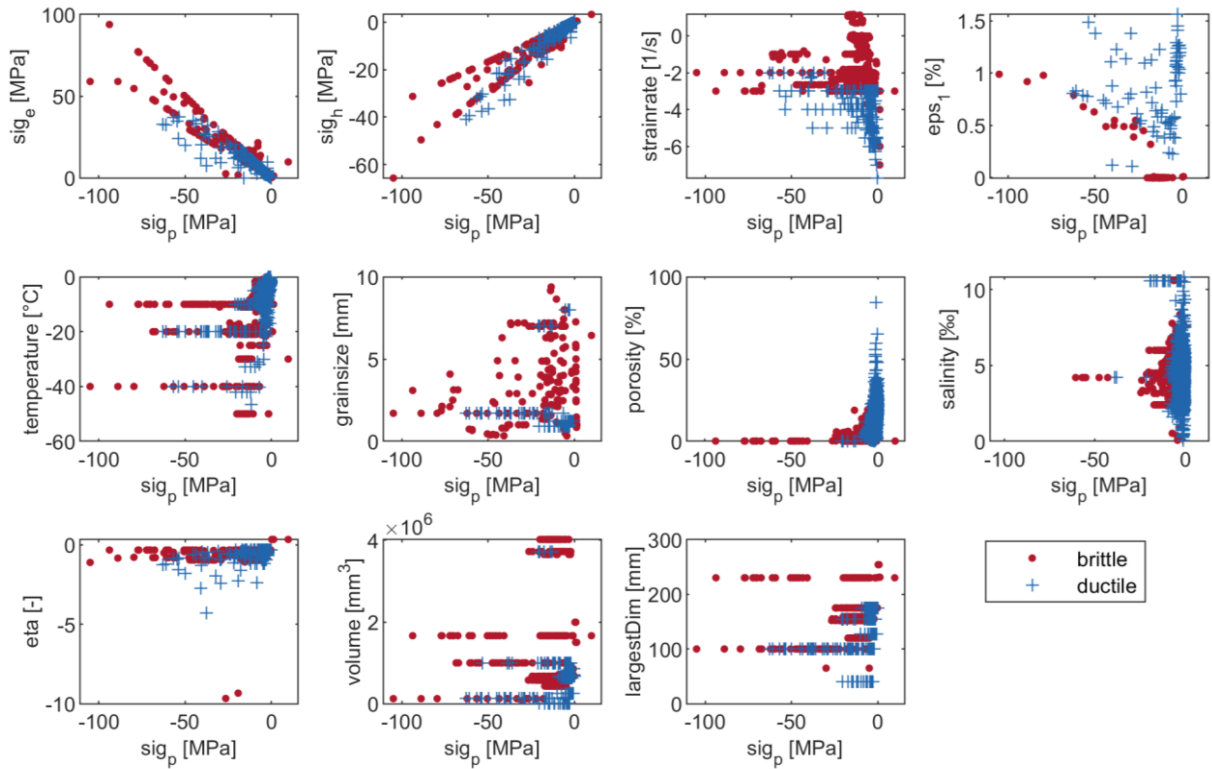

**Figure 4** - Matrix of scatter plots correlating each input feature with peak stress, grouped in ductile and brittle. The abbreviations are: sig_e = equivalent stress; sig_h = hydrostatic stress; eps_1 = limit strain; eta = triaxiality, η; largestD = largest dimension.

The significance of PCCs increases with the completeness of data. For instance, grain size is only documented for ~18% of the ductile observations. Therefore, the PCC for peak stress and grain size for this group is less significant.

**Table 2** - Input data for correlation analysis and completeness of observations. Low numbers of observations (< 75%) are highlighted in grey.

| Input: $X \in \mathbb{R}^{n \times m}$ | Completeness of observations [%] | | | |
|---|---|---|---|---|
| Dimensions (m=12) | brittle (n=979) | ductile (n=1329) | Freshwater ice (n=729) | Saltwater ice (n=2207) |
| peak stress | 100.0 | 99.5 | 99.0 | 100.0 |
| equivalent stress (sig_e) | 100.0 | 99.5 | 99.0 | 100.0 |
| hydrostatic stress (sig_h) | 100.0 | 99.5 | 99.0 | 100.0 |
| strain rate | 100.0 | 100 | 95.7 | 100.0 |
| limit strain (eps_1) | 7.9 | 5.6 | 13.6 | 8.2 |
| temperature | 92.7 | 98.1 | 100.0 | 91.6 |
| grain size | 18.6 | 6.8 | 54.7 | 0.0 |
| porosity | 52.2 | 74.9 | 25.1 | 67.3 |
| salinity | 49.3 | 83.6 | - | 88.2 |
| volume | 99.5 | 100.0 | 96.6 | 100.0 |
| eta / triaxiality | 100.0 | 99.5 | 99.0 | 100.0 |
| largest dimension | 99.5 | 100.0 | 99.0 | 100.0 |

The PCC between the different features and peak stress is shown in Figure 6, grouped in ductile, brittle as well as fresh- and saltwater ice observations. Note that peak stresses are negative in compression. For example, since strain rates are negatively correlated with peak stress (brittle group), higher strain rate values lead to lower peak stress, but higher absolute stress values, see Figure 5. The stress features, $\sigma_e$ and $\sigma_h$, correlate almost linearly with the peak stress. This is expected and indicates a correct calculation of the PCC.

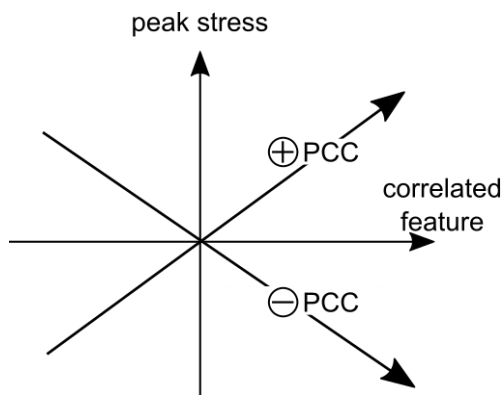

Figure 5 - PCC signs and peak stress

In the brittle group most features only mildly correlate with peak stress. The majority of PCCs is as expected and in line with relationships given in the literature. Higher temperature leads to lower absolute peak stress (Ince et al., 2016; Schulson, 1990), so do bigger grain size (Schulson, 1990, 1999) and porosity (ZhiJun et al., 2011). Higher strain rates lead to higher absolute peak stresses (Ince et al., 2016; Timco and Weeks, 2010), so do higher absolute triaxiality values (Schulson, 1999). In contrast, limit strain is strongly correlated. In displacement-controlled experiments, higher limit strain could indicate increasing time to failure and hence higher peak stress (Mellor and Cole, 1982). However, there is little data on limit strain, which renders conclusions uncertain. Salinity does not seem to be correlated. At least for brine volume this is in line with (Schulson, 1999). The correlations to geometry properties are inconclusive; the volume is not correlated whereas the largest dimension of the specimen is little correlated.

For the ductile group, strain rate is not correlated with peak stress, which is in contrast to other research (Kuehn and Schulson, 1994; Schulson and Buck, 1995; Snyder, 2015). However, these researchers also used data sets that were smaller and more balanced with respect to strain rate. Here, the input data is biased since the majority of strain rate observations (~75%) is within $10^{-3} < \dot{\epsilon} \leq 10^{-2}$. Limit strain does not appear to be correlated, but the significance of the PCC is low due to insufficient data. Grain size does not correlate, which is expected for common grain sizes (Goldsby and Kohlstedt, 1997; Schulson, 1999), but there is also only little data. Temperature correlates with peak stress, which is for example in accordance with (Kuehn and Schulson, 1994; Schulson, 1999). Porosity strongly correlates in comparison to other features, which is expected, see e.g. (Schulson and Gratz, 1999; Wang et al., 2018). Salinity is little correlated with peak stress. Although it is closely related to brine volume and porosity, which in turn affect the strength of ice, it does not seem significant. Triaxiality is more correlated, which is expected since higher confinement typically leads to higher stresses. Similar to the brittle group, correlations to geometry properties are inconclusive. The volume is not correlated whereas the largest dimension of the specimen is correlated as well as the largest dimension of the specimen.

The freshwater ice PCCs are similar to the brittle PCCs. This is probably due to the fact that the majority of freshwater ice observations are brittle (395 brittle; 142 ductile; 192 other/not defined). Most of the values are as expected and mildly correlated. Notably, limit strain and grain size are not correlated. For limit strain the incomplete data could distort the picture, but this does not apply to grain size where some correlation is expected.

The composition of saltwater ice behavior is: 1187 ductile, 584 brittle and 436 other/not defined. The PCCs give weight to strain rate, temperature and porosity, most other features are mildly correlated. As with the other groups, salinity is not correlated, and porosity seems to be a better indicator of peak stress. The saltwater ice group is the only group where both volume and largest dimension are mildly correlated, but with opposite signs.

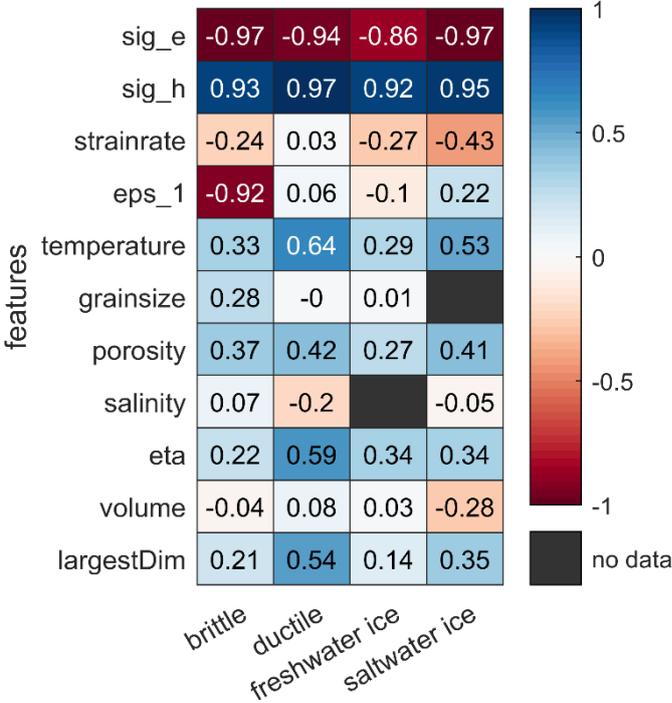

Figure 6 - PCC, between different features and peak stress, grouped in ductile, brittle, fresh- and saltwater ice. 1 is total positive linear correlation, 0 is no linear correlation, and –1 is total negative linear correlation. Values rounded to two decimals. The abbreviations are: sig_e = equivalent stress; sig_h = hydrostatic stress; eps_1 = failure strain; eta = triaxiality, η; largestD = largest dimension.

Overall, most PCCs are as expected and in line with relationships found in the literature. However, some PCCs are low or based on incomplete or imbalanced data. A few PCCs are inconclusive. For example, the features of volume and largest dimension may not be sufficient to capture the influence of geometry on peak stress. Additionally, the correlations change depending on the observed group. This emphasizes focusing on specific data sets rather than taking the full database as an input. Models that aim to calculate peak stress may need to consider different features, depending on the expected behavior.

### 4.3. Principal component analysis to identify irrelevant or redundant data

The principal component analysis (PCA) is applied to the following data sets: the whole set, ductile and brittle. The input features are similar to those used for the Pearson correlation coefficient (PCC), i.e. only numerical, non-categorical data. However, limit strain, grain size, porosity and salinity are excluded since the data for these features is incomplete. In contrast to the PCC analysis, missing values are crucial for the PCA since the data is processed as a whole matrix instead of pairs of features.

Therefore, if a feature is missing for an observation, the whole observation (i.e. the whole row of the data matrix) is removed. There exist many methods to deal with missing values (Ilin and Raiko, 2010), but this is not the focus of the current work. The applied basic PCA is suitable for complete observations only. After discarding the incomplete features, the remaining input features are; peak stress, strain rate, temperature, hydrostatic stress, equivalent stress, triaxiality, volume and largest dimension. Accordingly, the input is $X \in \mathbb{R}^{n \times m}$, with $m = 8$ features and $n(\text{all; ductile; brittle}) = 1861; 1193; 668$.

The data set is also normalized before applying the PCA. Original dimensions that intrinsically have a much larger variance will have more impact on the PCA. Hence, z-scoring (zero mean, unit standard deviation) is applied to normalize the feature columns. Below, only results for the whole set are presented. Grouped results are similar and can be found in the appendix.

The results comprise two points. First, the principal components (PC); these are the new base vectors which point in the direction of maximum variance in the data set. They can be represented as combinations of the original base vectors or features, respectively. This representation indicates which features introduce most variance to the data. Second, the variance in the data covered by the principal components. This indicates how many of the new base vectors would be needed to explain the variance in the data, which is shown in in Figure 7. The y-axis indicates the normalized cumulative variance and the x-axis the principal components. If all features except one were constant, one PC would explain all variance in the data. Thus in this case at least four principal components are needed to explain 90 % of variance in the data.

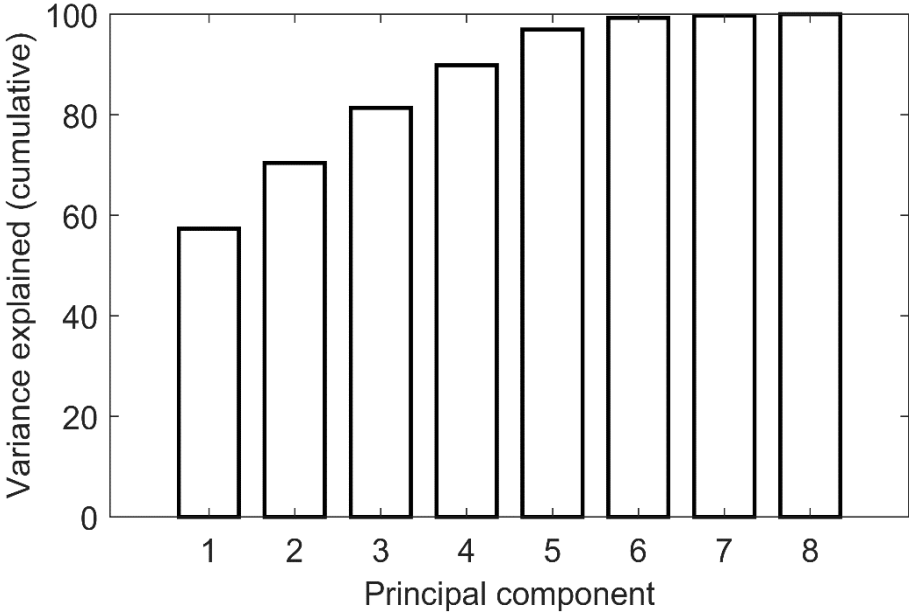

Figure 7 – Cumulative percentage of variance in the data explained by principal components for the complete data set.

The composition of the principal components in terms of the contribution of the original features is shown in Figure 8. There is no dominant feature in any of the first principal components. This indicates that the original features are of similar variance.

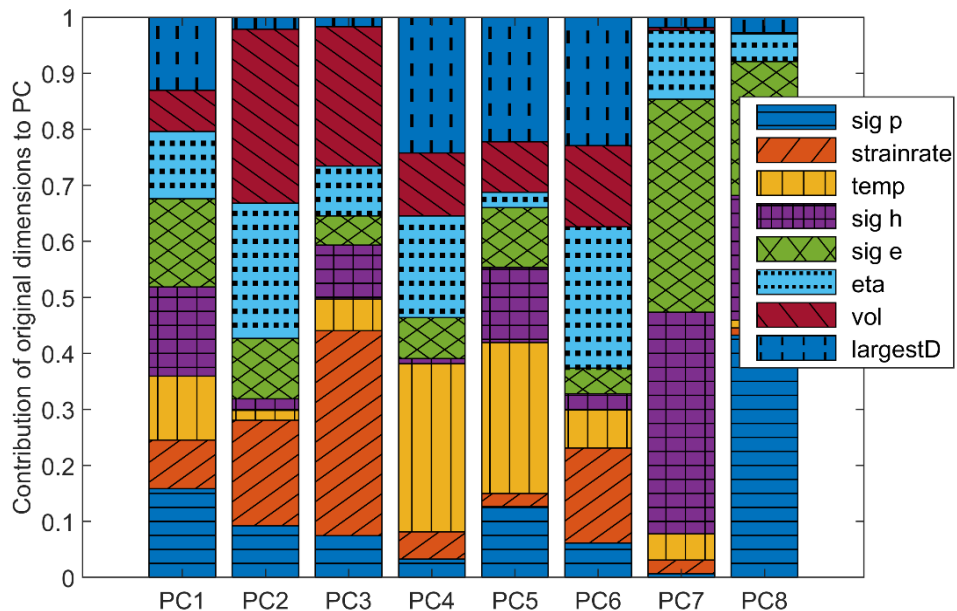

Figure 8 - Normalized contribution of original features to the principal components for the complete data set. Large contributions indicate principle components aligned parallel to the original feature dimension and thus large variance. The abbreviations are: sig_p = peak stress, temp = temperature, sig_h = hydrostatic stress; sig_e = equivalent stress; eta = triaxiality; vol = volume; largestD = largest dimension.

To summarize, neither dominant nor redundant or irrelevant features are found by the PCA, at least not for the current data set. It seems that the underlying relationship between the predictors (the influencing features) and response (peak stress) is not connected to the predictors' variability. Focusing on variance does not identify irrelevant or redundant features. This is not surprising considering the results of the correlation analysis via the Pearson's correlation coefficient.

### 4.4. How to predict behavior class with decision trees

Decision trees (DT) are a machine learning tool employed to predict the behavior type based on the input features described before. The input also comprises categorical features such as water type, which is in contrast to the Pearson correlation coefficient and the principal components analysis. Preliminary runs show that the DTs yield almost completely different results for fresh and saltwater ice. Hence the input data is split into saltwater and freshwater ice. The categorical feature type of columnar loading is excluded since this would require excluding granular ice. Peak stress and limit strain are excluded since in reality these figures are not known in advance. Lastly, water type is not considered since the differentiation in fresh and saltwater ice makes this feature redundant. The final input features are given in Table 3, for the completeness of these features see Table 7 in the appendix.

Table 3 - Input features for greedy backward elimination and decision trees

| **Saltwater ice DT** $X \in \{\mathbb{R}, categorical\}^{1773 \times 10}$ | **Freshwater ice DT** $X \in \{\mathbb{R}, categorical\}^{535 \times 9}$ |
|---|---|
| <ul><li>type of test</li><li>global strain rate</li><li>ice temperature</li><li>grain size</li></ul> | <ul><li>type of test</li><li>global strain rate</li><li>ice temperature</li><li>grain size</li></ul> |
| ▪ porosity<br>▪ salinity<br>▪ type of ice<br>▪ triaxiality<br>▪ volume<br>▪ largest dimension | ▪ porosity<br>▪ type of ice<br>▪ triaxiality<br>▪ volume<br>▪ largest dimension |

The quality of classification of a DT is evaluated based on the confusion matrix $C$ as shown in Table 4. It relates predicted to observed classes, e.g. the number of true positives (**TP**) indicates correctly predicted brittle observations. The most straightforward quality measure is accuracy, i.e. the percentage of correct predictions on the test data set, which is

$$ACC = \frac{TP + TN}{TP + FP + FN + TN} \qquad (7)$$

In the greedy backward elimination scheme (GBE), Matthews correlation coefficient (MCC) is used, (Chicco, 2017; Matthews, 1975), because we want to penalize wrong classifications and cannot assure balanced data sets. The MCC score is high only if the classifier is doing well on both possible outcome classes, i.e. negative and positive. It takes the following values:

$$MCC \begin{cases} +1; & \text{complete agreement} \\ 0; & \text{prediction no better than random} \\ -1; & \text{complete disagreement} \end{cases} \qquad (8)$$

And is calculated as:

$$MCC = \frac{TP \cdot TN - FP \cdot FN}{\sqrt{(TP + FP)(TP + FN)(TN + FP)(TN + FN)}} \qquad (9)$$

For the final training of the DTs, different quality measures are given. Cross validation is always applied. For more details on the different quality measures and the general application of DTs, the reader is referred to the appendix, Section 8.1.

**Table 4 - Confusion matrix for binary classification**

|  |  | observed / target class | |
|---|---|---|---|
|  |  | positive / brittle | negative / ductile |
| predicted / output class | positive / brittle | true positive - **TP** | false positive - **FP** |
| | negative / ductile | false negative - **FN** | true negative - **TN** |

### 4.4.1. Feature selection with greedy backward elimination

Five greedy backward elimination (GBE) runs are done for both ice types. The input data is as given in Table 3. Due to inherent randomness in the cross-validation process, the results can vary from run to run.

A typical result of the GBE for saltwater ice is shown in Figure 9. The x-axis shows the order in which features were eliminated from left to right. The y-axis indicates the quality of the decision tree (DT) in predicting behavior type and how this changes by eliminating features. The further right a feature is, the more helpful it is considered by the algorithm in predicting behavior type. For saltwater ice, different runs almost always yield strain rate and temperature as the two most important features. Salinity, type of test and triaxiality tend to be eliminated early. The order of the remaining features varies greatly from run to run due to the method-inherent random seeding. However, there is no feature that does not appear in the top five features at least once. Moreover, the elimination order barely affects the overall quality of the intermediate DTs. The MCC stays around 0.5 which is a mediocre value. It appears that after considering temperature and strain rate, the use of other features is interchangeable. Using less information (fewer features) does not lead to better or worse predictions.

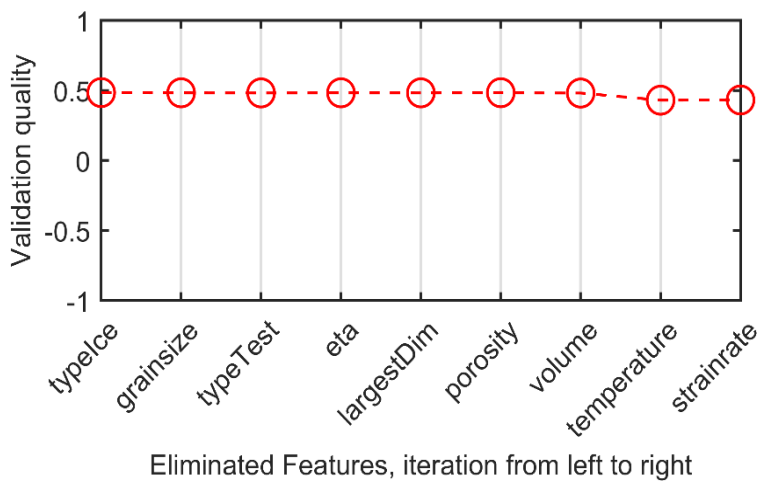

**Figure 9 - Example of greedy backwards elimination for salt water ice. Quality evaluated with Matthews correlation coefficient (MCC), -1 worst, +1 best.**

For freshwater ice, a typical result of the GBE is shown in Figure 10. In this case, different runs always yield strain rate as the most important feature and all other features appear interchangeable. Type of test, temperature and porosity tend to be eliminated early. Again, there is no feature that is not among the top five features during the five runs and the change in quality of intermediate trees is small. Compared to the MCC achieved for saltwater ice, the DT classification quality is significantly higher for freshwater ice. All in all, regarding the input data for DT training, strain rate should be used for freshwater ice whereas temperature and strain rate should be used for saltwater ice. Beyond that, the results of the feature selection are inconclusive. It appears that besides temperature and strain rate, most other features have some predictive value, but no feature dominates. It should be noted that this is only the case for the present choice of sub data sets and the prediction of behavior type.

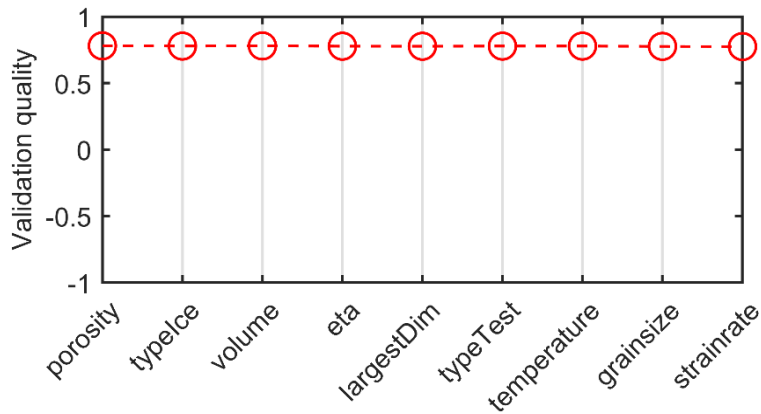

**Figure 10 - Example of greedy backwards elimination for fresh water ice. Quality evaluated with Matthews correlation coefficient (MCC), -1 worst, +1 best.**

### 4.4.2. Training of the final decision trees and results

The input data for training the final decision tree (DT) is again split into saltwater and freshwater ice. Based on the previous results, all features are used for training and the input data is the same as for the greedy backward elimination (Table 3). In addition to cross validation, the quality of the final DTs is also evaluated by resubstitution. That is, training and testing data set are the same.

An example for a saltwater ice DT is shown in Figure 11. Triangles represent decisions on the feature shown above the triangle. The criterion for going to the left branch is on the left side of the triangle and vice versa. As suggested by the feature selection process, strain rate and temperature are considered the best behavior predictors in the training of the final DT. The right branch seems redundant. This is true regarding classification since it ends up in the same class. However, the two leaves predict ductile behavior with different probabilities. The results are considered reasonable with respect to ice mechanics since ice tends to ductile behavior for lower strain rates and higher temperatures.

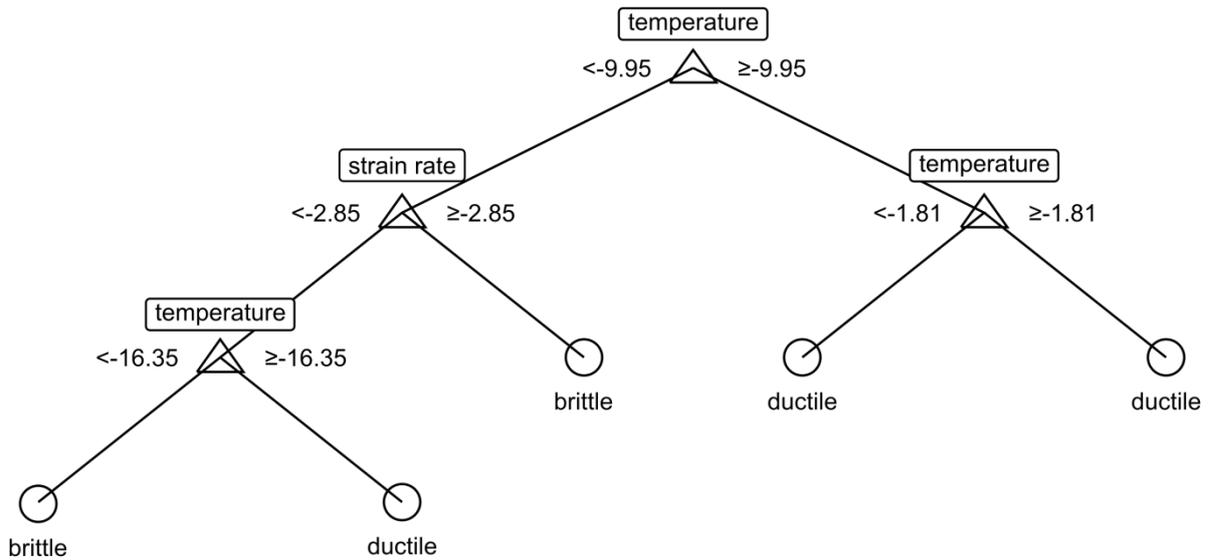

Figure 11 - Example of a final saltwater decision tree. All values rounded to two decimals. Strain rate values are logarithmized to base 10.

The confusion matrix $C$ is shown in Figure 12. The lowest cell on the right, $C(3,3)$, is the overall accuracy, 78.2 %, which does not seem bad at first glance. The last row and column indicate row- and column wise quality measures, e.g. $C(1,3) = TP/(TP + FP)$. The prediction of brittle behavior is good, with 8.8 % of all brittle observations misclassified. The prediction of ductile behavior is inaccurate with almost every fourth ductile observation (23.9 %) misclassified. The majority of misclassified observations is due to ductile misclassification $\left(\frac{FN}{FN+FP} = \frac{366}{366+21} = 94.6\ \%\right)$. This is also reflected by the quality measures given in saltwater ice is less predictable than brittle behavior.

Table 5. F1 score and MCC are more balanced measures than accuracy and consider the distribution of classes. As a result, their values are low compared to accuracy. It appears that ductile behavior for

|  | Accuracy (0 worst, 1 best) | F1 score (0, 1) | MCC (-1, 1) |
|---|---|---|---|
| **k-fold cross validation** | 0.780 ± 0.011 (k=10) | 0.506 ± 0.026 (k=10) | 0.488 ± 0.045 (k=10) |
| **resubstitution** | 0.780 | 0.522 | 0.484 |

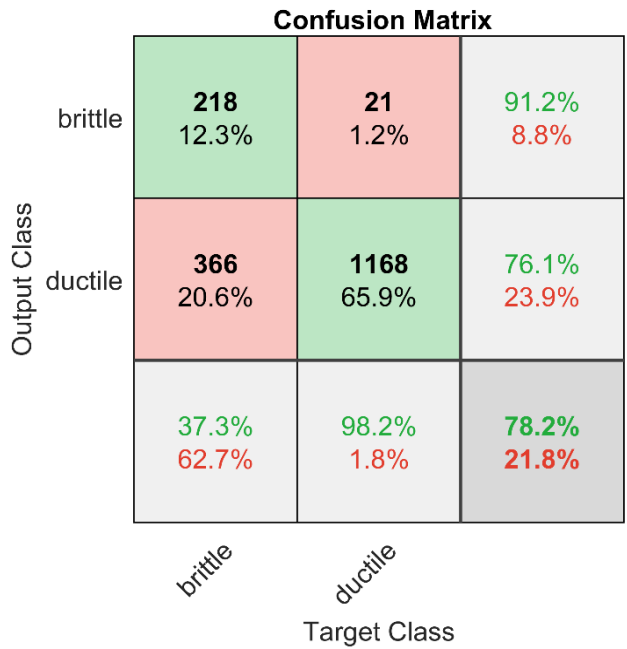

Figure 12 - Confusion matrix for the final salt water ice decision tree.

Table 5 - Quality measures for the final saltwater decision tree. Averaged from cross validation with standard deviation, for hold out and for resubstitution.

|  | Accuracy (0 worst, 1 best) | F1 score (0, 1) | MCC (-1, 1) |
| --- | --- | --- | --- |
| k-fold cross validation | 0.780 ± 0.011 (k=10) | 0.506 ± 0.026 (k=10) | 0.488 ± 0.045 (k=10) |
| resubstitution | 0.780 | 0.522 | 0.484 |

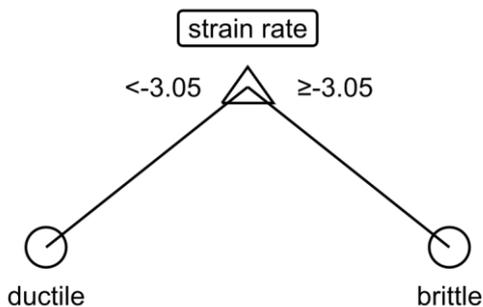

Figure 13 - Example of a final fresh water ice decision tree. Values rounded to two decimals and logarithmized to base 10.

An example of a freshwater ice DT is shown in Figure 13. Although all features are used as training data, the training proposes a decision tree which only considers strain rate as a predictor with a threshold strain rate of $\sim 10^{-3}$. Both the confusion matrix, Figure 14, and the quality measures given in Table 6, indicate a high quality of prediction. This confirms the results of the feature selection. It should be kept in mind that the data set for training this DT is smaller than for the saltwater ice DT (1773 to 535).

Both DTs are robust to changes of input data in terms of included features and hyper parameters, such as the maximum number of node splits. However, only the freshwater ice DT is also robust regarding the amount of data used for training. With just 30% of observations as input, the training of the freshwater ice DT already converges to the structure shown in Figure 13. The saltwater ice DT identifies temperature and strain rate as decisive features with little input data, but only converges to the structure shown in Figure 11 after using about 80% of the observations as input. This indicates that the behavior of saltwater ice is elusive, particularly in comparison to that of freshwater ice.

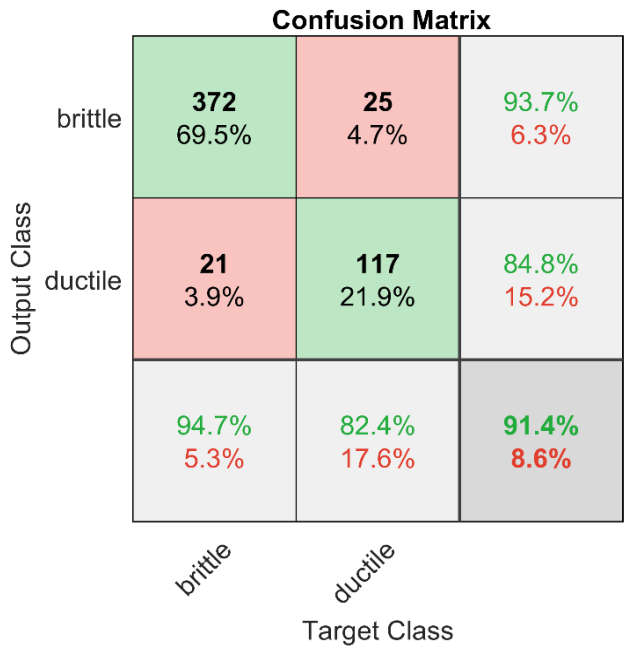

Figure 14 - Confusion matrix for the final fresh water decision tree.

Table 6 - Quality measures for the final freshwater decision tree. 1st row: averaged from cross validation with standard deviation. 2nd row: for resubstitution.

|  | Accuracy (0 worst, 1 best) | F1 score (0, 1) | MCC (-1, 1) |
|---|---|---|---|
| k-fold cross validation | 0.912 ± 0.037 (k=10) | 0.941 ± 0.025 (k=10) | 0.773 ± 0.100 (k=10) |
| resubstitution | 0.914 | 0.942 | 0.778 |

The results can be compared to the literature by looking at the ductile to brittle transition, which has been shown to depend on different features, such as strain rate, temperature, confinement, grain size and prestrain (Batto and Schulson, 1993; Renshaw et al., 2014; Schulson and Buck, 1995; Snyder et al., 2016). A commonly given transition strain rate for freshwater and saltwater ice is $\dot{\epsilon} \cong 10^{-4}$ and $\dot{\epsilon} \cong 10^{-3}$, respectively, at $-10°$ C, e.g. (Snyder et al., 2016). A surprising result is that the transition strain rate for freshwater ice given by the DT is one magnitude higher than the literature values ($\dot{\epsilon} \cong 10^{-3.05}$). A possible reason for this deviation is that the DT strain rate is based on a data set which lumps together the results of different experimental setups and conditions. Another unexpected result is that the DT transition strain rates for saltwater and freshwater ice are almost identical. However, for the saltwater ice DT, the transition strain rate only applies to temperatures lower than $-9.95°$ C. Hence it cannot be seen as a general transition strain rate in a strict sense. As such, a direct comparison to literature values or the freshwater transition strain rate is less meaningful. It should also be kept in mind that the saltwater ice DT misclassified about 22% of observations.

In all, among the mentioned features, the DTs identify temperature and strain rate as the most important ones. Prestrain is not included in the data yet. In general, the results are in line with the literature. The freshwater transition strain rate is a notable exception.

## 5. Discussion

To begin with, the focus is on the first part, the database and its composition (Section 2.2). It has to be kept in mind that the available data is partially incomplete. Nevertheless, it is straightforward to apply analysis methods to the database and the database is easily extendable in its current form. Also, the suggested list of features to be included (Section 2.1) is supported by the results. Neither the principal component analysis (PCA) nor the Pearson correlation coefficient (PCC) analysis unambiguously identify redundant or irrelevant features, but they point out trends. The greedy backwards elimination (GBE) scheme consistently gives weight to strain rate and temperature but is also inconclusive in ruling out other features. Hence, although the suggested list of features may not be complete, it does not include any unnecessary non-influential features. As such, the list is in line with the literature review in Section 2.1. Consequently, data collected in the future should include measurements of all these features wherever possible.

For the second part, the data analysis, it is emphasized that results can be biased by the input data and the way tools are applied. All results of the analysis are only valid for small scale experimental data. Furthermore all features are given from a global perspective, e.g. global strain rate or globally ductile behavior. Consequently, any results of the data analysis should only be used within this frame.

With respect to the most influential features on peak stress, few simple, one-dimensional relationships are found. Moreover, these relationships change depending on the input data, e.g. groups, number of observations, gaps etc. If sufficient data exists, and for a careful choice of input data, most PCCs indicate similar relationships as in the literature. In addition to that, the PCCs can put such relationships on a broad base of data and facilitate decisions on the inclusion of features in models. On the other hand, looking at the variance in the data does not result in the identification of irrelevant and redundant features. However, this can only be concluded from the current database and PCA approach. Different techniques of normalization or treatment of missing values, as well as larger and less imbalanced input data may lead to more conclusive results. Imbalance is apparent in many features, for instance strain rate, where 80% of observations are within a range $10^{-3} \leq \dot{\epsilon} \leq 10^{-2}$.

In terms of the question of behavior prediction in dependence of features, the GBE scheme and the DTs prove useful. Though, after several GBE runs, finally all features are used as input data for the DT training, the final DTs give clear recommendations on the most important features, i.e. temperature and strain rate. The predictive value of these two features is sufficient to predict ductile or brittle behavior with good accuracy. Particularly for freshwater ice the prediction model is simple yet accurate. Only the prediction of ductile behavior for saltwater ice seems more difficult to determine. The presented final DTs can be used in determining or verifying the global, qualitative behavior of ice models with only those two features. However, they are not readily applicable to making decisions on local behavior, e.g. on the element level of a material model.

Finally, though the methods applied to peak stress and behavior under load are different, it is noteworthy that they don't fully agree on the most important features. For freshwater ice, the final DT only uses strain rate as a predictor, whereas the PCCs suggest additionally using temperature, porosity and triaxiality. Of those four, the strain rate has the lowest correlation coefficient. For saltwater ice, the final DT is more in line with the PCCs but ignores porosity as the most correlated feature. However, this could be due to insufficient data. To sum up, the picture is not complete yet, but different features might be important, depending on the goal of peak stress or behavior prediction.

## 6. Conclusions and outlook

In essence, the results stress multi-dimensional dependencies between peak stress and behavior on one side and influential features and ice properties on the other. The one-dimensional relation between strain rate and ductile/brittle behavior for freshwater ice is a notable exception to this. This supports the use of large data sets to understand and capture ice behavior. A complete publication of experimental data in digital form is necessary to establish such large data sets, but seldom done, a notable exception is for instance (Strub-Klein and Sudom, 2012). The current database is available through the Institute of Ship Structural Design and Analysis at the Hamburg University of Technology. Other researchers are invited to use and extend the database. Furthermore, it is desirable to increase standardization in publishing data. In this regard, the list in Section 2.1 is a suggestion and open to discussion and revision.

Regarding the data analysis, it can be concluded that the chosen methods are suitable. The findings provide orientation on decisions which parameters to include in models and how to include them. The results also encourage expanding the toolbox of data analysis for material modeling with machine learning and statistical tools. The proposed workflow is generic and can be adapted to build DTs for other purposes. This is done by changing training data, target variables and hyper parameters. DTs can also be improved by executing automated optimization of hyper parameters. As mentioned, the use of the DTs in this work is briefly discussed along with detailed application parameters in the appendix, see Section 8.1. The current code is open source.

A conceivable option would be to apply the methodology to related areas, such as full-scale data for ice-structure interaction. As especially ice-induced vibrations are not fully understood, it is desirable to collect data for all possibly influential features, e.g. ice properties or metocean (Nord et al., 2016). In this case, instead of focusing on the material model input, the goal could be to identify general features that dominate the interaction process. It would also be valuable to know how much data is actually necessary to conclude trends.

## 7. Acknowledgements

We would like to thank the German Federal Ministry for Economic Affairs and Energy (BMWi) for funding this research under the project reference number 0324022B. Likewise, the financial support of the German Research Foundation (DFG) under the grant EH485/1-1 is gratefully acknowledged. It is stated that the BMWi and DFG are not responsible for any of the content of this publication.

We would also like to thank Torodd Nord for valuable and helpful discussions.

**Declaration of interest**: none.

## 8. Appendix

### 8.1. Structure of the database

The current database is an excel table, where rows correspond to observations and columns to features and their units. The features correspond to those described in Section 2.2. The first two columns are reserved for comments on the calculation of principal stresses, experiment numbers etc. For example, sometimes one principal stress is given along with the relation of this stress to the other principal stresses, i.e. $R$ – values, for a definition e.g. (Renshaw and Schulson, 2001). In that case it can make sense to include the $R$ – values. The first two columns after the observations are reserved for additional comments and the source of the data. Simple excel filtering can be used to extract specific data sets and their sources.

The observations are sorted top to bottom in the following order; uni-/bi-/triaxial compression, shear, uniaxial tension. Within those classes, they are sorted in ascending order by year of the publication of the source, i.e. oldest publications first and most actual last. Most comment and source cells within the table are combined cells which span all observations taken from that source.

## 8.2. Application of decision trees

From a machine learning perspective, the problem at hand is a supervised, binary classification task. As a classifier we use a decision tree (DT) with Gini-index based splitting up to ten times. The curvature test is used as a split predictor since the DT deals with continuous and categorical data (Loh and Shih, 1997). The minimum leaf size is set to 50 observations. This is done to avoid that leaves only represent one data set in the whole database. In that case, the split that leads to the leaf could be biased by the way that specific data was obtained, e.g. measurement errors, specific test setups, etc. Nevertheless, the DTs are robust against changes of leaf size in the range 10 to 100. The leaf size affects the number of decisions (nodes), but the decisions for the first nodes as shown in the results, Section 4, remain the same. Moreover, surrogate decision splits are used when the value of the optimal split predictor is missing. As a result, the DT is less reliant on complete data. Lastly, the hyper parameters (leaf size, maximum number of splits) were not optimized.

To compute the quality of the model, stratified k-fold cross validation is employed. This means that the data is partitioned into k nearly equally sized folds. Then, k iterations of training and validation are carried out. Within each iteration, k-1 folds are used for training and the remaining fold is used for validation (Refaeilzadeh et al., 2009). A 10-fold cross validation is used because it appears to be a good "rule of thumb" value (Chicco, 2017). Furthermore, the folds are stratified, i.e. balanced with regards to classes. After ensuring the model quality in the building phase of the model with cross validation, the final DT is trained on the whole data. The underlying assumption is that the surrogate models, which are trained on the partitioned data, are equivalent to the whole data model. In other words, if the cross validation was sucessful, it is allowable to train the final DT on the complete data.

In order to access the classification quality, a quality measure has to be chosen. Typically, the results computed by a supervised learner are collected in the confusion matrix. For a binary classification problem, the confusion matrix contains the number of i) true positives (TP), ii) true negatives (TN), iii) false positives (FP) and iv) false negatives (FN), see Section 4.4. Classic quality measures are accuracy , true positive rate ('recall') $TPR$ and F1 score:

$$ACC = \frac{TP + TN}{TP + TN + FP + FN} \tag{10}$$

$$TPR = \frac{TP}{TP + FN} \tag{11}$$

$$F_1 = \frac{2TP}{2TP + FP + FN} \tag{12}$$

However, these measures are not sensitive to false negatives and only valid for balanced data sets. A data set is considered balanced if the labels are distributed equally for all observations (here: 50% brittle, 50% ductile). As we want to penalize wrong classifications and cannot assure balanced datasets, Matthews correlation coefficient MCC is used as classification quality measure (Chicco, 2017; Jurman et al., 2012; Matthews, 1975).

$$MCC = \frac{TP \cdot TN - FP \cdot FN}{\sqrt{(TP + FP)(TP + FN)(TN + FP)(TN + FN)}} \tag{9}$$

The MCC score is high only if the classifier is doing well on both possible outcome classes, i.e. negative and positive. If any of the parentheses in the denominator are zero, MCC is set to zero. It can be shown that MCC is always zero if the denominator is zero.

### 8.3. Additional diagrams and tables

In Table 7, a general list of features is given including the completeness of data. This list comprises all features which are used as input at the beginning of the data processing. In comparison to the original database, the following features are excluded; Peak stress as originally described by author and type of limit strain (4. and 8. in the list of Section 2.2). Furthermore, for the data analysis, the 2nd and 3rd principal limit strains are not used, since almost no data is included.

Table 7 - General features and completeness, grouped. Low numbers of observations (< 75%) are highlighted in grey.

|  | Completeness of observations [%]. | | | | |
| --- | --- | --- | --- | --- | --- |
| **Features** | all (n=2939) | saltwater ice (n=2207) | freshwater ice (n=729) | brittle (n=979) | ductile (n=1329) |
| typeTest | 100.0 | 100.0 | 100.0 | 100.0 | 100.0 |
| sig_1 | 99.8 | 100.0 | 99.0 | 100.0 | 99.5 |
| sig_2 | 99.8 | 100.0 | 99.0 | 100.0 | 99.5 |
| sig_3 | 99.8 | 100.0 | 99.0 | 100.0 | 99.5 |
| peakStress | 99.8 | 100.0 | 99.0 | 100.0 | 99.5 |
| typeBehavior | 83.7 | 83.5 | 84.6 | - | - |
| strainrate | 98.9 | 100.0 | 95.7 | 100.0 | 100.0 |
| eps_1 | 9.5 | 8.2 | 13.6 | 7.9 | 5.6 |
| eps_2 | 1.9 | 1.9 | 0.0 | 5.8 | 0.0 |
| eps_3 | 2.1 | 2.8 | 0.0 | 6.2 | 0.0 |
| temperature | 93.7 | 91.6 | 100.0 | 92.7 | 98.1 |
| grainsize | 13.6 | 0.0 | 54.7 | 18.6 | 6.8 |
| porosity | 56.8 | 67.3 | 25.1 | 52.2 | 74.9 |
| salinity | 66.2 | 88.2 | - | 49.3 | 83.6 |
| typeIce | 99.7 | 99.7 | 100.0 | 100.0 | 99.5 |
| typeWater | 99.9 | - | - | 100.0 | 100.0 |
| typeColumnarLoading[5] | 95.2 | 99.3 | 74.2 | 92.3 | 99.1 |
| sig_h | 99.8 | 100.0 | 99.0 | 100.0 | 99.5 |
| sig_e | 99.8 | 100.0 | 99.0 | 100.0 | 99.5 |
| eta | 99.8 | 100.0 | 99.0 | 100.0 | 99.5 |
| width[6] | 100.0 | 100.0 | 100.0 | 100.0 | 100.0 |
| depth | 100.0 | 100.0 | 100.0 | 100.0 | 100.0 |
| length | 100.0 | 100.0 | 100.0 | 100.0 | 100.0 |
| diameter[7] | 100.0 | 100.0 | 100.0 | 100.0 | 100.0 |
| volume | 99.0 | 100.0 | 96.6 | 99.5 | 100.0 |
| largestDim[8] | 100.0 | 100.0 | 100.0 | 100.0 | 100.0 |

---

[5] Calculated in percentage of columnar ice observations.
[6] Calculated in percentage of cuboid specimens. Same for depth and length.
[7] Calculated in percentage of cylindrical specimens.
[8] Not in the original database. Always calculated if geometry data is available.

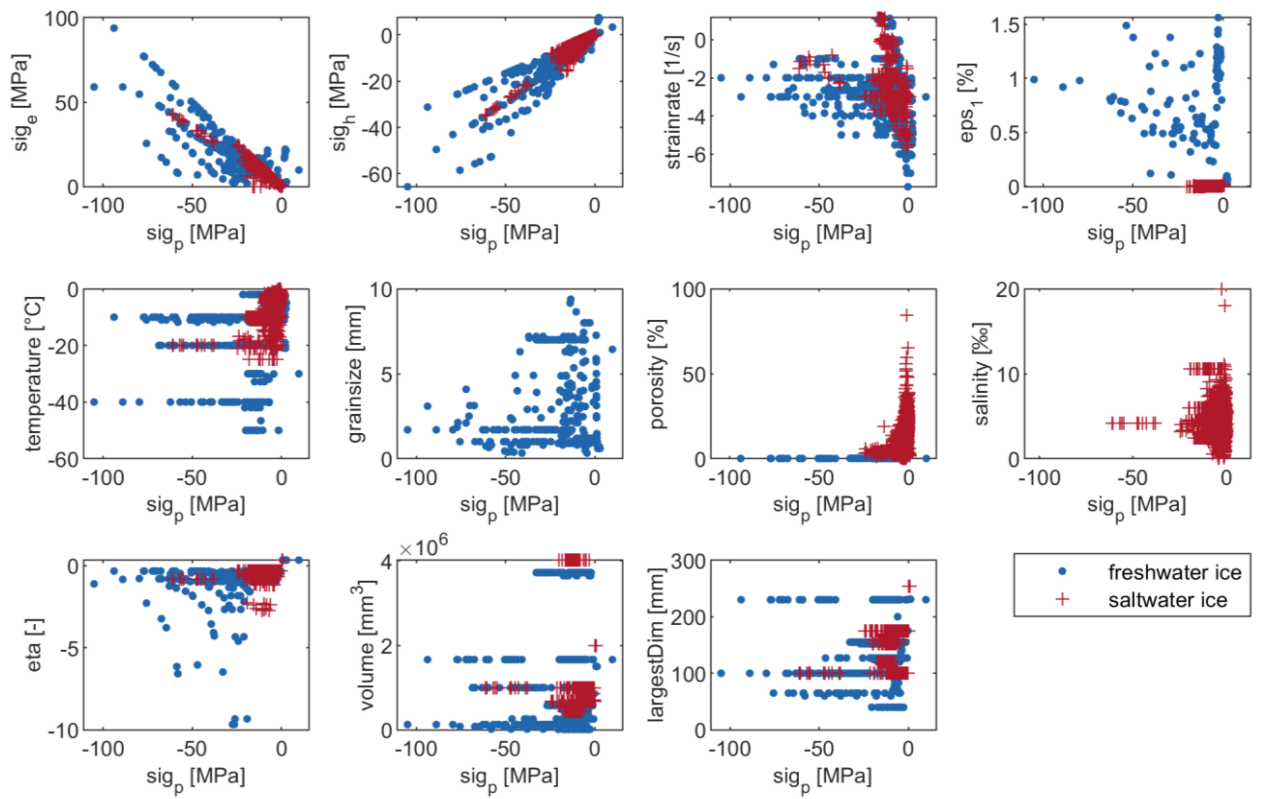

Figure 15 - Matrix of scatter plots correlating each input feature with peak stress, grouped in freshwater and saltwater ice.

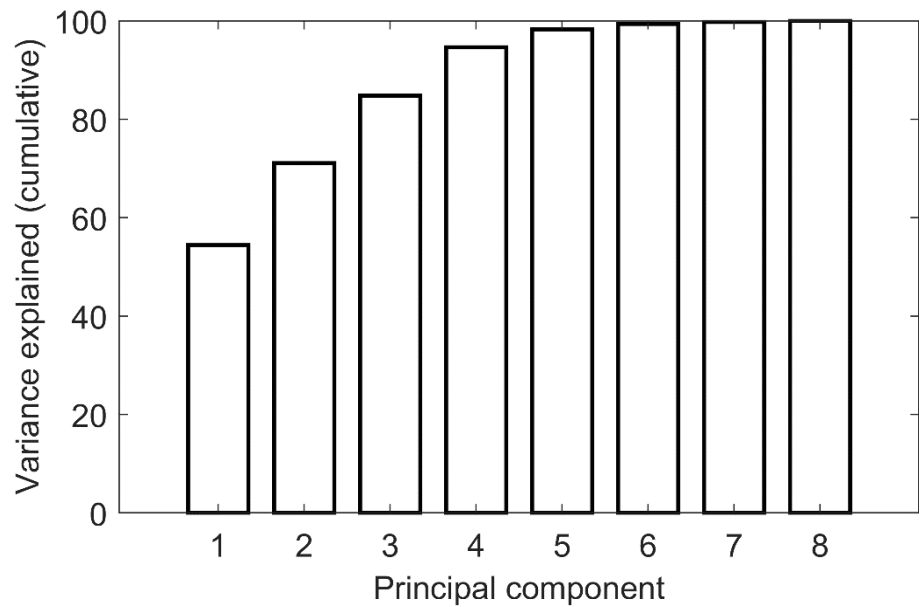

Figure 16 - Cumulative percentage of variance in the data explained by principal components for brittle observations.

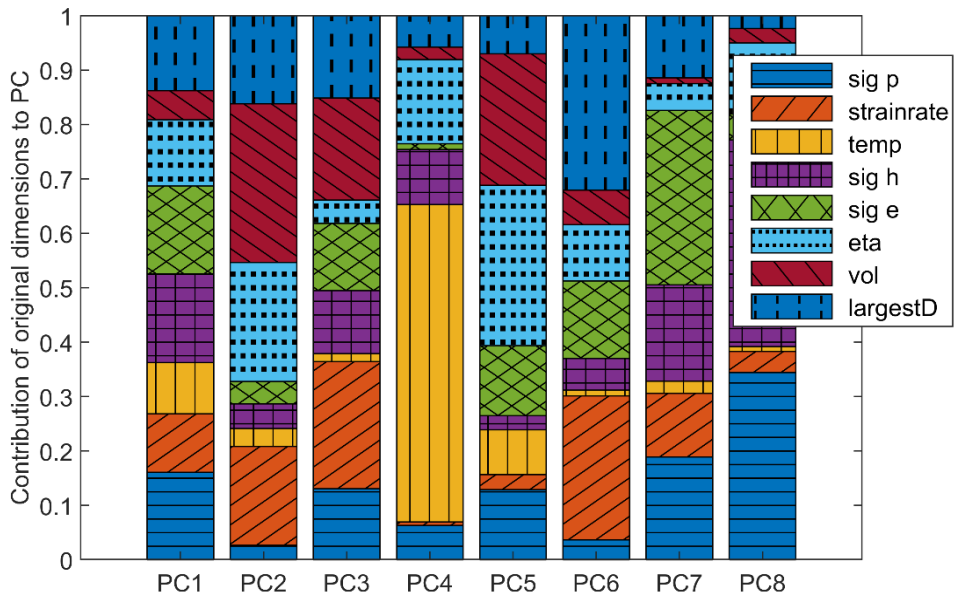

Figure 17 - Normalized contribution of original features to principal components for the brittle observations. Large contributions indicate principle components aligned parallel to the original feature dimension and thus large variance. The abbreviations are: sig_p = peak stress, temp = temperature, sig_h = hydrostatic stress; sig_e = equivalent stress; eta = triaxiality; vol = volume; largestD = largest dimension.

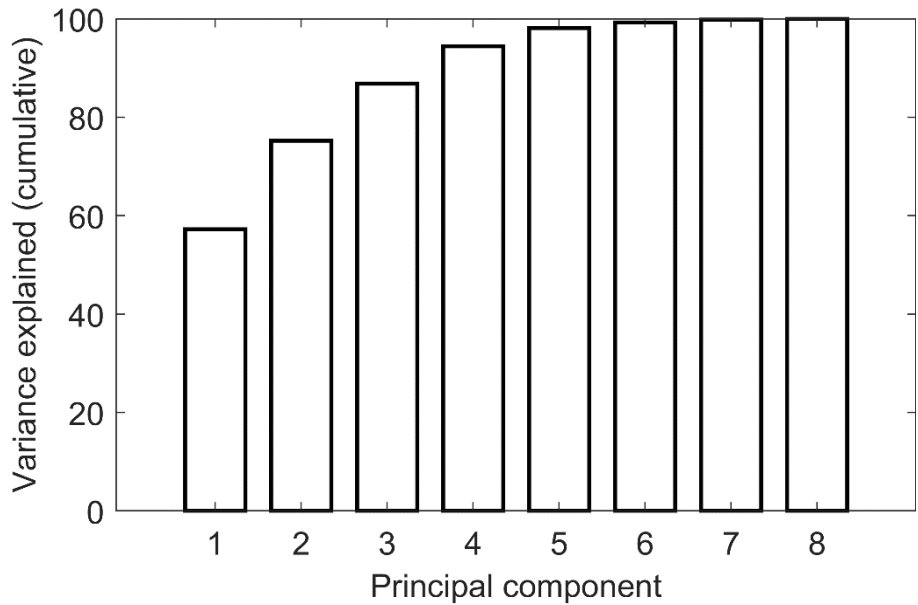

Figure 18 - Cumulative percentage of variance in the data explained by principal components for ductile observations.

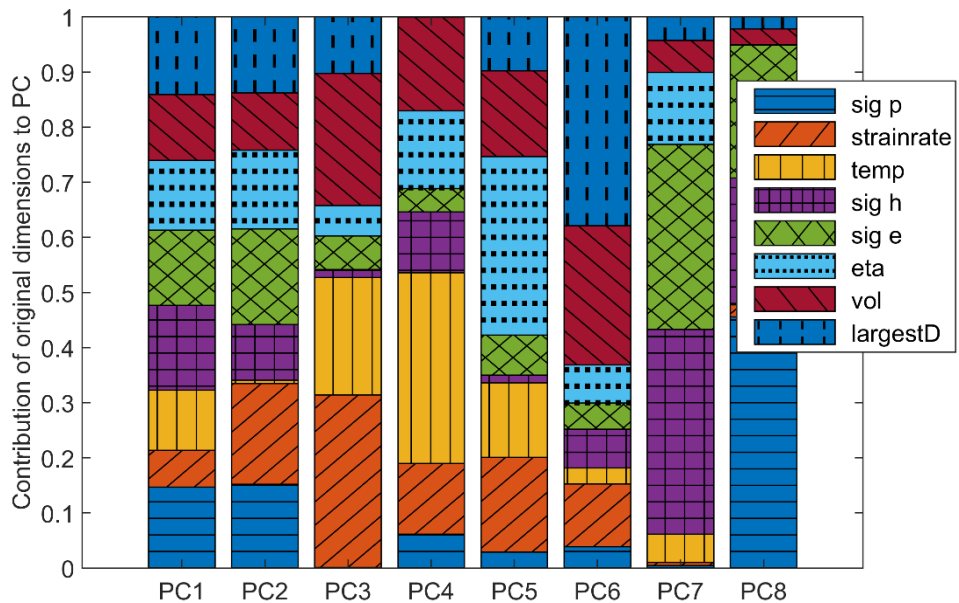

Figure 19 - Normalized contribution of original features to principal components for the ductile observations. Large contributions indicate principle components aligned parallel to the original feature dimension and thus large variance. The abbreviations are: sig_p = peak stress, temp = temperature, sig_h = hydrostatic stress; sig_e = equivalent stress; eta = triaxiality; vol = volume; largestD = largest dimension.

### 8.4. Database and matlab codes

If you would like to obtain the database or contribute with experimental data, please contact Sören Ehlers, ehlers@tuhh.de. Institute for Ship Structural Design and Analysis. Am Schwarzenberg Campus 4 C. 21073 Hamburg. Germany.

The matlab codes are licensed under the GNU General Public License (gnu.org/licenses/gpl-2.0.html), if you would like to obtain them, please contact Leon Kellner, leon.kellner@tuhh.de.

## References

Arakawa, M., Maeno, N., 1997. Mechanical strength of polycrystalline ice under uniaxial compression. Cold Regions Science and Technology 26 (3), 215–229. doi:10.1016/S0165-232X(97)00018-9.

Batto, R., Schulson, E., 1993. On the ductile-to-brittle transition in ice under compression. Acta Metallurgica et Materialia 41 (7), 2219–2225. doi:10.1016/0956-7151(93)90391-5.

Chicco, D., 2017. Ten quick tips for machine learning in computational biology. BioData Mining 10 (1), 205. doi:10.1186/s13040-017-0155-3.

Currier, J., Schulson, E., 1982. The tensile strength of ice as a function of grain size. Acta Metallurgica 30 (8), 1511–1514. doi:10.1016/0001-6160(82)90171-7.

Dempsey, J., Adamson, R., Mulmule, S., 1999. Scale effects on the in-situ tensile strength and fracture of ice: Part II: First-year sea ice at Resolute, N.W.T. Int J Fract 95 (1/4), 347–366. doi:10.1023/A:1018650303385.

Dempsey, J., Cole, D., Wang, S., 2018. Tensile fracture of a single crack in first-year sea ice. Philosophical transactions. Series A, Mathematical, physical, and engineering sciences 376 (2129). doi:10.1098/rsta.2017.0346.

Effelsberg, J., Haufe, A., Feucht, M., Neukamm, F., Du Bois, P., 2012. On parameter identification for the GISSMO damage model, in: 12th International LS-DYNA Users Conference. Metal Forming.

Ehlers, S., Polojärvi, A., Vredeveldt, A., Quinton, B., Kim, E., Ralph, F., Sirkar, J., Moslet, P.O., Fukui, T., Kuehnlein, W., Wan, Z., 2018. Committee V.6 Arctic Technology, in: Proceedings of the 20th International Ship and Offshore Structures Congress (ISSC 2018), Liege, Belgium; Egmond an Zee, Netherlands;

Fletcher, N., 1970. The chemical physics of ice. Cambridge U.P, London.

Fulcher, B., Jones, N., 2014. Highly Comparative Feature-Based Time-Series Classification. IEEE Trans. Knowl. Data Eng. 26 (12), 3026–3037. doi:10.1109/TKDE.2014.2316504.

Gibbons, J., 1986. Nonparametric Statistical Inference. 2nd. Ed. Statistics: Textbooks and monographs vol. 65. Biom. J. 28 (8), 936. doi:10.1002/bimj.4710280806.

Golding, N., Schulson, E., Renshaw, C., 2010. Shear faulting and localized heating in ice: The influence of confinement. Acta Materialia 58 (15), 5043–5056. doi:10.1016/j.actamat.2010.05.040.

Golding, N., Schulson, E., Renshaw, C., 2012. Shear localization in ice: Mechanical response and microstructural evolution of P-faulting. Acta Materialia 60 (8), 3616–3631. doi:10.1016/j.actamat.2012.02.051.

Golding, N., Snyder, S.A., Schulson, E., Renshaw, C., 2014. Plastic faulting in saltwater ice. J. Glaciol. 60 (221), 447–452. doi:10.3189/2014JoG13J178.

Goldsby, D., Kohlstedt, D., 1997. Grain boundary sliding in fine-grained Ice I. Scripta Materialia 37 (9), 1399–1406. doi:10.1016/S1359-6462(97)00246-7.

Gratz, E., Schulson, E., 1997. Brittle failure of columnar saline ice under triaxial compression. J. Geophys. Res. 102 (B3), 5091–5107. doi:10.1029/96JB03738.

Grennerat, F., Montagnat, M., Castelnau, O., Vacher, P., Moulinec, H., Suquet, P., Duval, P., 2012. Experimental characterization of the intragranular strain field in columnar ice during transient creep. Acta Materialia 60 (8), 3655–3666. doi:10.1016/j.actamat.2012.03.025.

Härdle, W., Simar, L., 2015. Applied Multivariate Statistical Analysis. Springer Berlin Heidelberg, Berlin, Heidelberg.

Hastie, T., Tibshirani, R., Friedman, J., 2017. The elements of statistical learning: Data mining, inference, and prediction, Second edition, corrected at 12th printing 2017 ed. Springer, New York, NY, 745 pp.

Häusler, F., 1981. Dreidimensionales Bruchkriterium für Meer-Eis: Bericht Nr. E 113/81. Hamburgische Schiffbau-Versuchsanstalt.

Hawkes, I., Mellor, M., 1972. Deformation and Fracture of Ice Under Uniaxial Stress. J. Glaciol. 11 (61), 103–131. doi:10.1017/S002214300002253X.

Haynes, F., 1978. Effect of temperature on the strength of snow ice: CRREL Report 78-27. CRREL, Hanover.

Haynes, F., Mellor, M., 1977. Measuring the uniaxial compressive strength of ice, in: International Glaciological Society (Ed.), Journal of Glaciology, vol. 19, pp. 213–223.

Hobbs, P., 2010. Ice physics. Oxford Univ. Press, Oxford, 837 pp.

Høyland, K., 2007. Morphology and small-scale strength of ridges in the North-western Barents Sea. Cold Regions Science and Technology 48 (3), 169–187. doi:10.1016/j.coldregions.2007.01.006.

Iliescu, D., Schulson, E., 2004. The brittle compressive failure of fresh-water columnar ice loaded biaxially. Acta Materialia 52 (20), 5723–5735. doi:10.1016/j.actamat.2004.07.027.

Ilin, A., Raiko, T., 2010. Practical Approaches to Principal Component Analysis in the Presence of Missing Values. Journal of Machine Learning Research 11, 1957–2000.

Ince, S., Kumar, A., Paik, J., 2016. A new constitutive equation on ice materials. Ships and Offshore Structures, 1–14. doi:10.1080/17445302.2016.1190122.

Jones, S., 1982. The Confined Compressive Strength of Polycrystalline Ice. J. Glaciol. 28 (98), 171–178. doi:10.1017/S0022143000011874.

Jones, S., 1997. High Strain-Rate Compression Tests on Ice. J. Phys. Chem.

Jordaan, I., 2001. Mechanics of ice–structure interaction. Engineering Fracture Mechanics 68 (17-18), 1923–1960. doi:10.1016/S0013-7944(01)00032-7.

Jurman, G., Riccadonna, S., Furlanello, C., 2012. A Comparison of MCC and CEN Error Measures in Multi-Class Prediction. PLoS ONE 7 (8), e41882. doi:10.1371/journal.pone.0041882.

Keeley, R., McDonald, R., 2015. Part III: Principal component analysis: bridging the gap between strain, sex and drug effects. Behavioural brain research 288, 153–161. doi:10.1016/j.bbr.2015.03.027.

Kuehn, G., Schulson, E., 1994. The mechanical properties of saline ice under uniaxial compression, in: , Annals of Glaciology, vol. 19, pp. 39–48.

Larrañaga, P., Calvo, B., Santana, R., Bielza, C., Galdiano, J., Inza, I., Lozano, J., Armañanzas, R., Santafé, G., Pérez, A., Robles, V., 2006. Machine learning in bioinformatics. Briefings in Bioinformatics 7 (1), 86–112. doi:10.1093/bib/bbk007.

Lehto, P., Remes, H., Saukkonen, T., Hänninen, H., Romanoff, J., 2014. Influence of grain size distribution on the Hall–Petch relationship of welded structural steel. Materials Science and Engineering: A 592, 28–39. doi:10.1016/j.msea.2013.10.094.

Liu, A., Wang, L., Li, H., Gong, J., Liu, X., 2017. Correlation Between Posttraumatic Growth and Posttraumatic Stress Disorder Symptoms Based on Pearson Correlation Coefficient: A Meta-Analysis. The Journal of nervous and mental disease 205 (5), 380–389. doi:10.1097/NMD.0000000000000605.

Loh, W., Shih, Y., 1997. Split Selection Methods for Classification Trees, in: Statistica Sinica, vol. 7, pp. 815–840.

Määttänen, M., 2015. Ice induced frequency lock-in vibrations - converging towards consensus, in: Proceedings of the 23th International Port and Ocean Engineering under Arctic Conditions. Port and Ocean Engineering under Arctic Conditions, Trondheim.

Mantovani, G., Lamontagne, M., Varin, D., Cerulli, G., Beaulé, P., 2012. Comparison of total hip arthroplasty surgical approaches by Principal Component Analysis. Journal of biomechanics 45 (12), 2109–2115. doi:10.1016/j.jbiomech.2012.05.041.

Matthews, B.W., 1975. Comparison of the predicted and observed secondary structure of T4 phage lysozyme. Biochimica et Biophysica Acta (BBA) - Protein Structure 405 (2), 442–451. doi:10.1016/0005-2795(75)90109-9.

Mellor, M., Cole, D., 1982. Deformation and failure of ice under constant stress or constant strain-rate. Cold Regions Science and Technology 5 (3), 201–219. doi:10.1016/0165-232X(82)90015-5.

Mizuno, Y., 1998. Effect of Hydrostatic Confining Pressure on the Failure Mode and Compressive Strength of Polycrystalline Ice. J. Phys. Chem. B 102 (2), 376–381. doi:10.1021/jp963163b.

Montagnat, M., Castelnau, O., Bons, P., Faria, S., Gagliardini, O., Gillet-Chaulet, F., Grennerat, F., Griera, A., Lebensohn, R., Moulinec, H., Roessiger, J., Suquet, P., 2014. Multiscale modeling of ice deformation behavior. Journal of Structural Geology 61, 78–108. doi:10.1016/j.jsg.2013.05.002.

Moslet, P.O., 2007. Field testing of uniaxial compression strength of columnar sea ice. Cold Regions Science and Technology 48 (1), 1–14. doi:10.1016/j.coldregions.2006.08.025.

Nadreau, J., Nawwar, A., Wang, Y., 1991. Triaxial Testing of Freshwater Ice at Low Confining Pressures. J. Offshore Mech. Arct. Eng 113 (3), 260. doi:10.1115/1.2919929.

Nord, T., Øiseth, O., Lourens, E., 2016. Ice force identification on the Nordströmsgrund lighthouse. Computers & Structures 169, 24–39. doi:10.1016/j.compstruc.2016.02.016.

Palmer, A., 2008. Dimensional analysis and intelligent experimentation. World Scientific, New Jersey, 154 pp.

Palmer, A., Dempsey, J., 2009. Model Tests in Ice, in: Proceedings of the 20[th] International Conference on Port and Ocean Engineering under Arctic Conditions, Luleå, Sweden.

Palmer, A., Croasdale, K., 2013. Arctic offshore engineering. World Scientific, Hackensack New Jersey, xiv, 357 pages.

Petrovic, J., 2003. Review Mechanical properties of ice and snow. Journal of Materials Science 38 (1), 1–6. doi:10.1023/A:1021134128038.

Ralph, F., Jordaan, I., 2017. Local design pressures during ship ram events modeling the ocurrence and intensity of high pressure zones, in: Proceedings of the ASME 2017 36[th] International Conference on Ocean, Offshore and Arctic Engineering, Trondheim.

Refaeilzadeh, P., Tang, L., Huan, L., 2009. Cross-Validation, in: Liu, L., Özsu, M. (Eds.), Encyclopedia of Database Systems. Springer US, Boston, MA, pp. 532–537.

Renshaw, C., Schulson, E., 2001. Universal behaviour in compressive failure of brittle materials. Nature 412 (6850), 897–900. doi:10.1038/35091045.

Renshaw, C., Golding, N., Schulson, E., 2014. Maps for brittle and brittle-like failure in ice. Cold Regions Science and Technology 97, 1–6. doi:10.1016/j.coldregions.2013.09.008.

Renshaw, C., Schulson, E., 2017. Strength-limiting mechanisms in high-confinement brittle-like failure: Adiabatic transformational faulting. J. Geophys. Res. Solid Earth 122 (2), 1088–1106. doi:10.1002/2016JB013407.

Richter-Menge, J., Jones, K., 1993. The tensile strength of first-year sea ice. J. Glaciol. 39 (133), 609–618. doi:10.1017/S0022143000016506.

Rist, M., Murrell, S., 1994. Ice triaxial deformation and fracture. J. Glaciol. 40 (135), 305–318. doi:10.1017/S0022143000007395.

Roebuck, B., Phatak, C., Birks-Agnew, I., 2004. A Comparison of the Linear Intercept and A Comparison of the Linear Intercept and Equivalent Circle Methods for Grain Size Measurement in WC/Co Hardmetals, Teddington, Middlesex, UK.

Sanderson, T., 1988. Ice Mechanics and Risks to Offshore Structures, 1., st Edition. Softcover version of original hardcover edition 1987 ed. Springer Netherland, Berlin.

Sankari, E., Manimegalai, D., 2017. Predicting membrane protein types using various decision tree classifiers based on various modes of general PseAAC for imbalanced datasets. Journal of theoretical biology 435, 208–217. doi:10.1016/j.jtbi.2017.09.018.

Schulson, E., Lim, P., Lee, R., 1984. A brittle to ductile transition in ice under tension. Philosophical Magazine A 49 (3), 353–363. doi:10.1080/01418618408233279.

Schulson, E., 1990. The brittle compressive fracture of ice. Acta Metallurgica et Materialia 38 (10), 1963–1976. doi:10.1016/0956-7151(90)90308-4.

Schulson, E., Buck, S., 1995. The ductile-to-brittle transition and ductile failure envelopes of orthotropic ice under biaxial compression. Acta Metallurgica et Materialia 43 (10), 3661–3668. doi:10.1016/0956-7151(95)90149-3.

Schulson, E., 1999. The structure and mechanical behavior of ice. JOM 51 (2), 21–27. doi:10.1007/s11837-999-0206-4.

Schulson, E., Gratz, E., 1999. The brittle compressive failure of orthotropic ice under triaxial loading. Acta Materialia 47 (3), 745–755. doi:10.1016/S1359-6454(98)00410-8.

Schulson, E., Fortt, A., Iliescu, D., Renshaw, C., 2006. Failure envelope of first-year Arctic sea ice: The role of friction in compressive fracture. J. Geophys. Res. 111 (C11). doi:10.1029/2005JC003235.

Schulson, E., Duval, P., 2009. Creep and fracture of ice. Cambridge University Press, Cambridge.

Schulson, E., 2015. Low-speed friction and brittle compressive failure of ice: fundamental processes in ice mechanics. International Materials Reviews 60 (8), 451–478. doi:10.1179/1743280415Y.0000000010.

Schulson, E., 2018. Friction of sea ice. Philosophical transactions. Series A, Mathematical, physical, and engineering sciences 376 (2129). doi:10.1098/rsta.2017.0336.

Smith, T., Schulson, E., 1993. The brittle compressive failure of fresh-water columnar ice under biaxial loading. Acta Metallurgica et Materialia 41 (1), 153–163. doi:10.1016/0956-7151(93)90347-U.

Snyder, S., 2015. Mechanical behavior and elastic properties of prestrained columnar ice. Ph.D., Hanover, New Hampshire.

Snyder, S., Schulson, E., Renshaw, C., 2016. Effects of prestrain on the ductile-to-brittle transition of ice. Acta Materialia 108, 110–127. doi:10.1016/j.actamat.2016.01.062.

Stone, B., Jordaan, I., Xiao, J., Jones, S., 1997. Experiments on the damage process in ice under compressive states of stress. Journal of Glaciology 43 (143).

Strub-Klein, L., Sudom, D., 2012. A comprehensive analysis of the morphology of first-year sea ice ridges. Cold Regions Science and Technology 82, 94–109. doi:10.1016/j.coldregions.2012.05.014.

Strub-Klein, L., 2017. A Statistical Analysis of First-Year Level Ice Uniaxial Compressive Strength in the Svalbard Area. J. Offshore Mech. Arct. Eng 139 (1), 11503. doi:10.1115/1.4034526.

Tayefi, M., Saberi-Karimian, M., Esmaeili, H., Zadeh, A., Ebrahimi, M., Mohebati, M., Heidari-Bakavoli, A., Azarpajouh, M., Heshmati, M., Safarian, M., Nematy, M., Parizadeh, S., Ferns, G., Ghayour-Mobarhan, M., 2018. Evaluating of associated risk factors of metabolic syndrome by using decision tree. Comp Clin Pathol 27 (1), 215–223. doi:10.1007/s00580-017-2580-6.

Timco, G., Frederking, R., 1984. An investigation of the failure envelope of granular/discontinuous-columnar sea ice. Cold Regions Science and Technology 9 (1), 17–27. doi:10.1016/0165-232X(84)90044-2.

Timco, G., O'Brien, S., 1994. Flexural strength equation for sea ice. Cold Regions Science and Technology 22 (3), 285–298. doi:10.1016/0165-232X(94)90006-X.

Timco, G., Weeks, W., 2010. A review of the engineering properties of sea ice. Cold Regions Science and Technology 60 (2), 107–129. doi:10.1016/j.coldregions.2009.10.003.

von Bock und Polach, F., Molyneux, D., 2017. Model ice: a review of its capacity and identification of knowledge gaps, in: Proceedings of the ASME 2017 36[th] International Conference on Ocean, Offshore and Arctic Engineering. International Conference on Ocean, Offshore and Arctic Engineering, Trondheim.

Wang, Q., Li, Z., Lei, R., Lu, P., Han, H., 2018. Estimation of the uniaxial compressive strength of Arctic sea ice during melt season. Cold Regions Science and Technology 151, 9–18. doi:10.1016/j.coldregions.2018.03.002.

Weiss, J., Schulson, E., 1995. The failure of fresh-water granular ice under multiaxial compressive loading. Acta Metallurgica et Materialia 43 (6), 2303–2315. doi:10.1016/0956-7151(94)00421-8.

Yue, Q., Guo, F., Kärnä, T., 2009. Dynamic ice forces of slender vertical structures due to ice crushing. Cold Regions Science and Technology 56 (2-3), 77–83. doi:10.1016/j.coldregions.2008.11.008.

Zhang, L., Li, Z., Jia, Q., Li, G., Huang, W., 2011. Uniaxial Compressive Strengths of Artificial Freshwater Ice. AMR 243-249, 4634–4637. doi:10.4028/www.scientific.net/AMR.243-249.4634.

ZhiJun, L., LiMin, Z., Leppäranta, M., GuangWei, L., 2011. Experimental study on the effect of porosity on the uniaxial compressive strength of sea ice in Bohai Sea. Sci. China Technol. Sci. 54 (9), 2429–2436. doi:10.1007/s11431-011-4482-1.